\documentclass[11pt,a4paper]{article}

\usepackage{mlmodern}
\usepackage{amsfonts,amssymb,amsthm}
\usepackage{upgreek,mathrsfs}
\usepackage{bold-extra}
\usepackage{microtype}

\usepackage[dvipsnames]{xcolor}
\definecolor{maroon}{RGB}{167,74,74}
\definecolor{magreen}{RGB}{59,125,37}
\definecolor{mablue}{RGB}{59,136,195}
\definecolor{mayellow}{RGB}{242,147,24}
\colorlet{mabrown}{maroon!50!magreen}
\colorlet{maorange}{maroon!50!mayellow}
\colorlet{macyan}{magreen!50!mablue}

\colorlet{red}{maroon}
\colorlet{green}{magreen}
\colorlet{blue}{mablue}
\colorlet{yellow}{mayellow}
\colorlet{brown}{mabrown}
\colorlet{orange}{maorange}
\colorlet{cyan}{macyan}

\definecolor{rred}{RGB}{167,33,74}
\definecolor{bblue}{RGB}{29,136,255}
\definecolor{ppurple}{RGB}{113,84,165}
\definecolor{ppink}{RGB}{255,55,219}

\RequirePackage{hyperref}
\hypersetup{colorlinks,
    linkcolor=mablue,
    citecolor=magreen,
    urlcolor=maroon,
    anchorcolor=mablue,
    filecolor=mablue,
    menucolor=mablue}

\RequirePackage[top=12mm,bottom=12mm,left=30mm,right=30mm,head=12mm,includeheadfoot]{geometry}
\makeatletter
\g@addto@macro\bfseries{\boldmath}
\makeatother

\makeatletter
\renewcommand\section{\@startsection{section}{1}{\z@}%
  {-3.5ex \@plus -1.3ex \@minus -.7ex}%
  {2.3ex \@plus.4ex \@minus .4ex}%
  {\large\bfseries}}
\renewcommand\subsection{\@startsection{subsection}{2}{\z@}%
    {-2.3ex\@plus -1ex \@minus -.5ex}%
    {1.2ex \@plus .3ex \@minus .3ex}%
    {\normalsize\bfseries}}
\renewcommand\subsubsection{\@startsection{subsubsection}{3}{\z@}%
    {-2.3ex\@plus -1ex \@minus -.5ex}%
    {1ex \@plus .2ex \@minus .2ex}%
    {\normalsize\bfseries}}
\renewcommand\paragraph{\@startsection{paragraph}{4}{\z@}%
    {1.75ex \@plus1ex \@minus.2ex}%
    {-1em}%
    {\normalsize\bfseries}}
\renewcommand\subparagraph{\@startsection{subparagraph}{5}{\parindent}%
    {1.75ex \@plus1ex \@minus .2ex}%
    {-1em}%
    {\normalsize\bfseries}}
\makeatother

\RequirePackage{tocloft}

\RequirePackage[nottoc,notlot,notlof]{tocbibind}

\usepackage{comment}

\usepackage{import}
\usepackage{enumitem}
\usepackage{csquotes}

\RequirePackage[thicklines]{cancel}

\usepackage[new]{old-arrows}

\newcommand{\nn}{\nonumber \\}

\renewcommand{\geq}{\geqslant}
\renewcommand{\leq}{\leqslant}

\usepackage{physics}
\usepackage{stackengine}

\newcommand{\parti}{\mathcal{Z}} 
\newcommand{\pd}{\partial} 

\newcommand\ccolon{\text{\scalebox{1.2}{\raisebox{0ex}{\stackanchor[.28ex]{.}{.}}}}} 
\newcommand{\no}[1]{\; \ccolon {#1} \ccolon \;} 
\newcommand{\bm}[1]{\boldsymbol{#1}} 

\newcommand{\tps}[2]{\texorpdfstring{\ensuremath{#1}}{#2}} 

\newcommand{\suchthat}{\;\middle|\;} 

\renewcommand{\vec}{\boldsymbol} 

\newcommand{\w}{\mathbin{\scalebox{0.8}{$\wedge$}}} 
\newcommand{\ex}[1]{\mathrm{e}^{#1}} 
\newcommand{\ii}{i} 

\let\U\undefined

\newcommand{\R}{\mathbb{R}} 
\newcommand{\Z}{\mathbb{Z}} 
\renewcommand{\S}{S} 

\newcommand{\set}[1]{\qty{#1}} 

\newcommand{\U}{\mathrm{U}} 
\newcommand{\SO}{\mathrm{SO}} 
\newcommand{\alg}[1]{\mathfrak{#1}} 
\newcommand{\su}{\alg{su}} 
\newcommand{\so}{\alg{so}} 

\renewcommand{\t}[1]{{\text{#1}}} 
\newcommand{\xmapsto}[1]{\overset{#1}{\mapsto}} 

\makeatletter
\renewcommand{\xmapsto}[2][]{\ext@arrow 0599{\mapstofill@}{#1}{#2}}
\def\mapstofill@{\arrowfill@{\mapstochar\relbar}\relbar\rightarrow}
\makeatother

\def \cA {\mathcal{A}}
\def \cB {\mathcal{B}}

\def \cF {\mathcal{F}}

\def \cH {\mathcal{H}}

\def \cN {\mathcal{N}}
\def \cO {\mathcal{O}}

\def \cZ {\mathcal{Z}}

\DeclareSymbolFont{bbold}{U}{bbold}{m}{n}
\DeclareSymbolFontAlphabet{\mathbbold}{bbold}

\def \bb1 {{\mathbb{1}}}

\def \sfH {\mathsf{H}}

\def \sfq {\mathsf{q}}

\let\tilde\thintilde
\newcommand{\tilde}{\widetilde}

\usepackage{tikz} 

\usetikzlibrary{decorations.markings,arrows.meta}

\tikzset{line/.style={line width=0.25mm},
curve/.style={line,smooth,tension=1},
->-/.style={decoration={
  markings,
  mark=at position #1 with {\arrow[>=stealth]{>}}},postaction={decorate}},
-<-/.style={decoration={
  markings,
  mark=at position #1 with {\arrow[>=stealth]{<}}},postaction={decorate}},
}

\tikzset{bg/.style={opacity=.5}}

\usepackage{tikz-cd}

\RequirePackage{mdframed}
\usepackage{empheq}

\def\fdiffd{\mathrm{D}}
\DeclareDocumentCommand\fdifferential{ o g d() }{ 
    \IfNoValueTF{#2}{
        \IfNoValueTF{#3}
        {\fdiffd\IfNoValueTF{#1}{}{^{#1}}}
        {\mathinner{\fdiffd\IfNoValueTF{#1}{}{^{#1}}\argopen(#3\argclose)}}
    }
    {\mathinner{\fdiffd\IfNoValueTF{#1}{}{^{#1}}#2} \IfNoValueTF{#3}{}{(#3)}}
}
\DeclareDocumentCommand\DD{}{\fdifferential}

\DeclareDocumentCommand\variation{ o g d() }{ 
    \IfNoValueTF{#2}{
        \IfNoValueTF{#3}
        {\updelta \IfNoValueTF{#1}{}{^{#1}}}
        {\mathinner{\updelta \IfNoValueTF{#1}{}{^{#1}}\argopen(#3\argclose)}}
    }
    {\mathinner{\updelta \IfNoValueTF{#1}{}{^{#1}}#2} \IfNoValueTF{#3}{}{(#3)}}
}

\def\sdiffd{\mathbf{d}}
\DeclareDocumentCommand\sdifferential{ o g d() }{ 
    \IfNoValueTF{#2}{
        \IfNoValueTF{#3}
        {\sdiffd\IfNoValueTF{#1}{}{^{#1}}}
        {\mathinner{\sdiffd\IfNoValueTF{#1}{}{^{#1}}\argopen(#3\argclose)}}
    }
    {\mathinner{\sdiffd\IfNoValueTF{#1}{}{^{#1}}#2} \IfNoValueTF{#3}{}{(#3)}}
}

\DeclareDocumentCommand\wedgecommutator{ l m m }{\braces#1{\lbrack}{\rbrack}{#2\w #3}} 

\DeclareMathOperator{\volume}{vol}
\DeclareDocumentCommand\vol{}{\opbraces{\volume}}

\DeclareMathOperator{\spanof}{span}
\DeclareDocumentCommand\spn{}{\opbraces{\spanof}}

\renewcommand{\ip}[2]{\left\langle#1,#2\right\rangle}

\RequirePackage{interval}
\intervalconfig{soft open fences}

\usepackage[noabbrev]{cleveref}

\crefname{subsection}{subsection}{subsections}
\crefname{equation}{}{}

\numberwithin{equation}{section}

\usepackage[style=numeric-comp, eprint=true, natbib, backend=biber, maxbibnames=5, giveninits=true, isbn=false, url=false, doi=false, date=year, sorting=none, useprefix=true]{biblatex}

\DeclareFieldFormat*{title}{\emph{#1}}

\DeclareFieldFormat*{journaltitle}{#1}

\DeclareFieldFormat{pages}{#1}

\DeclareFieldFormat{labelalpha}{\textsc{#1}}

\renewbibmacro{in:}{}

\renewbibmacro*{date}{\printtext[parens]{\printdate}}

\renewbibmacro*{issue+date}{%
    \printfield{issue}%
    \setunit*{\addspace}%
    \usebibmacro{date}%
    \newunit}

\renewbibmacro*{publisher+location+date}{%
    \printlist{location}%
    \iflistundef{publisher}
        {\setunit*{\addcomma\space}}
        {\setunit*{\addcolon\space}}%
    \printlist{publisher}%
    \setunit*{\addspace}%
    \usebibmacro{date}%
    \newunit}

\renewbibmacro*{institution+location+date}{%
    \printlist{location}%
    \iflistundef{institution}
        {\setunit*{\addcomma\space}}
        {\setunit*{\addcolon\space}}%
    \printlist{institution}%
    \setunit*{\addspace}%
    \usebibmacro{date}%
    \newunit}

\renewbibmacro*{organization+location+date}{%
    \printlist{location}%
    \iflistundef{organization}
        {\setunit*{\addcomma\space}}
        {\setunit*{\addcolon\space}}%
    \printlist{organization}%
    \setunit*{\addspace}%
    \usebibmacro{date}%
    \newunit}

\renewbibmacro*{location+date}{%
    \printlist{location}%
    \setunit*{\addspace}%
    \usebibmacro{date}%
    \newunit}

\AtEveryBibitem{\clearfield{urldate}}
\AtEveryBibitem{\clearfield{month}}
\AtEveryBibitem{\clearfield{day}}
\AtEveryBibitem{\clearfield{issn}}
\AtEveryBibitem{\clearfield{version}}

\DeclareFieldFormat{doilink}{%
	\iffieldundef{doi}{%
		\iffieldundef{url}{%
			\iffieldundef{isbn}{%
				\iffieldundef{issn}{%
					#1%
				}{%
					\href{http://books.google.com/books?vid=ISSN\thefield{issn}}{#1}%
				}%
			}{%
				\href{http://books.google.com/books?vid=ISBN\thefield{isbn}}{#1}%
			}%
		}{%
			\href{\thefield{url}}{#1}%
		}%
	}{%
		\href{http://dx.doi.org/\thefield{doi}}{#1}%
	}%
}

\DeclareBibliographyDriver{article}{%
	\usebibmacro{bibindex}%
	\usebibmacro{begentry}%
	\usebibmacro{author/translator+others}%
	\setunit{\labelnamepunct}\newblock
	\usebibmacro{title}\addcomma\space%
	\newunit
	\printlist{language}%
	\newunit\newblock
	\usebibmacro{byauthor}%
	\newunit\newblock
	\usebibmacro{bytranslator+others}%
	\newunit\newblock
	\printfield{version}%
	\newunit\newblock
	\printtext[doilink]{%
		\usebibmacro{journal+issuetitle}%
		\newunit
		\usebibmacro{byeditor+others}%
		\newunit
		\usebibmacro{note+pages}%
	}%
	\newunit\newblock
	\iftoggle{bbx:isbn}
	{\printfield{issn}}
	{}%
	\newunit\newblock
	\usebibmacro{doi+eprint+url}%
	\newunit\newblock
	\usebibmacro{addendum+pubstate}%
	\setunit{\bibpagerefpunct}\newblock
	\usebibmacro{pageref}%
	\usebibmacro{finentry}}

\newcommand{\ds}{\t{dS}} 
\newcommand{\rds}{R_{\ds}} 
\newcommand{\kip}[2]{\ip{#1}{#2}_\t{KG}} 
\newcommand{\lie}{\pounds}
\newcommand{\dvol}{\dd{\!\vol}}
\newcommand{\Vol}{\vol}
\usepackage{slashed}

\begin{document}

\begin{center}
	{\Large\textbf{Axions on de Sitter space}} \\
  \bigskip
	Vasileios A. Letsios,\kern-3pt$^{*}$ and Stathis Vitouladitis$^{\dagger}$ \\
	\bigskip
	\footnotesize{
    {\normalsize $^{*}$}
		Service de Physique de l' Univers, Champs et Gravitation, \\ UMONS, Place du Parc 20, 7000 Mons, BE \\[2ex]
		{\normalsize $^{\dagger}$}
		\textrm{Physique Théorique et Mathématique, Université Libre de Bruxelles \& \\ International Solvay Institutes, CP 231, 1050 Brussels, BE} \\
    \bigskip
		\href{mailto:vasileios.letsios@umons.ac.be}{\small \sf vasileios.letsios@umons.ac.be}, \quad
		\href{mailto:stathis.vitouladitis@ulb.be}{\small \sf stathis.vitouladitis@ulb.be}
	}
\end{center}

\bigskip

\begin{abstract}
\noindent We study a massless minimally coupled compact scalar, or axion, on global $D$-dimensional de~Sitter space (dS$_D$). From a Lorentzian perspective, we quantise the theory canonically, determine the quantum dS charges, and find that the axion zero mode supplies a quantum-mechanical factor beyond the oscillator Fock space $\cF_\t{osc}$. The full Hilbert space is $\cH=L^2(\S^1)\otimes\cF_\t{osc}$, with the integer quantum-mechanical momentum on $L^2(\S^1)$ identified with the conserved $\U(1)$ shift charge. The relevant single-particle unitary irreducible representation (UIR) of the dS group, $\mathrm{SO}(D,1)$, captures the oscillator sector, but misses the zero mode. We find that the neutral zero-particle state is \ds-invariant and normalisable. Charged zero-particle states are normalisable, but only $\mathrm{SO}(D)$ invariant. This implies that geodesic observers related by dS boosts do not agree on the number of particles in a charged sector, an effect absent in QFTs equipped only with the standard Bunch--Davies vacuum. We further compute field-strength Wightman two-point functions in charged sectors and find that they are Hadamard. For non-zero charge they are not \ds-invariant at finite global times, but they are asymptotically so at early and late times. We complement this analysis with a Euclidean perspective. The ordinary $D$-sphere path integral, $\parti_{S^D}$, written in terms of Harish-Chandra characters, has access only to the neutral sector. Charged sectors require vertex-operator insertions, and summing over them gives a decorated sphere path integral, $\widehat{\parti}_{S^D}=\parti_\t{QM}\,\parti_{S^D}$ that captures the entire Hilbert space, with $\parti_\t{QM}$ denoting the partition function of a quantum rotor at a dimension-dependent effective temperature. Finally, in dS$_3$, we use the duality between an axion and a photon to translate our results to electromagnetism. The oscillator sector relates to the photon UIR, while the axion zero mode gives rise to magnetic monopole sectors, which are allowed by the topology of global dS$_3$.
\end{abstract}

\pagebreak 

{\setlength{\parskip}{0pt}\tableofcontents}
\vspace{10pt}

\section{Introduction}\label{sec:intro}

The early universe is believed to have passed through a period of rapid accelerated expansion, known as inflation \cite{Baumann, Linde:1981mu}. The spacetime geometry of this epoch is approximated by de~Sitter ($\ds$) spacetime, the maximally symmetric solution of vacuum Einstein's equations with positive cosmological constant. Interestingly, recent observational data suggest that our universe is currently accelerating \cite{PlanckCollab, Suzuki, SDSS, Cole}, and might asymptote to another de~Sitter phase. Both eras call for a quantum theory on an expanding spacetime. A complete answer should come from quantum gravity in asymptotically de~Sitter spacetimes \cite{Witten,Anninos_musings}. As a step towards a more complete understanding of cosmologically relevant spacetimes, we study quantum field theory (QFT) on fixed \ds\ backgrounds.

In the present paper, we study the massless minimally coupled scalar on $D$-dimensional global de~Sitter spacetime, $\ds_D$. This theory has two main incarnations. In one, the scalar is \emph{non-compact} and takes values in $\R$, while in the other it is \emph{compact} and valued on a circle. The second variant is also called an axion. The non-compact case is well-studied \cite{Allen:1985ux,Allen:1987tz,Kirsten:1993ug,Gibbons-Higuchi, Gibbons-Higuchi, Tolley:2001gg, Anninos:2011kh}, already revealing interesting structures. By contrast, the compact case has only recently come under direct study \cite{Chakraborty:2023eoq, Chakraborty:2025mhh}. Given its simplicity, and its possible relevance for cosmological applications \cite{Freese:1990rb,Arkani-Hamed:2003xts}, it is somewhat surprising that a comprehensive account of its quantum theory on $\ds_D$ has not been available. We undertake this task here. We focus on the  quantum theory of the compact massless minimally coupled scalar on $\ds_D$ and we report a variety of interesting results.

To set the stage, we temporarily leave de~Sitter and define the compact scalar on a general $D$-dimensional spacetime, $X_D$. By compact scalar, also called an axion or axion-like field, we mean a map from $D$-dimensional spacetime, $X_D$, to a circle:
\begin{align}
    \phi:X_D \to \S^1~.
\end{align}
Equivalently, it is an ordinary real scalar field $\phi(x)$, together with the equivalence relation
\begin{align}\label{eq:compactness}
    \phi(x)\sim \phi(x)+2\pi~.
\end{align}
At the level of our analysis, $\phi$ can either be a scalar or a pseudoscalar. 
We conventionally take the scalar to be dimensionless, and the radius of the target-space circle to be 1. It participates in a quantum theory with action
\begin{align}\label{eq:action-compact}
    S[\phi] = \int \dd[D]{x}\,\sqrt{-g}\ \qty(-\frac{f^2}{2}\,g^{\mu\nu}\pd_\mu\phi\, \pd_\nu \phi - v(\phi))~,
\end{align}
in Lorentzian signature, with mostly-plus conventions. Here $v$ is a periodic potential, $v(\phi)=v(\phi+2\pi)$. The scalar may also couple to other fields of the theory. See, e.g., \cite{Arvanitaki:2009fg}. We will meet several such couplings below. The constant $f$, the ``axion decay constant'', carries dimensions of $L^{(2-D)/2}$, where $L$ is a characteristic length scale, for instance the \ds\ radius, $\rds$, in the present paper. It cannot be removed by a field redefinition, even when $v=0$, as it would re-enter in \cref{eq:compactness} and affect observables.

In the free case, $v=0$, the theory has an emergent global $\U(1)$ shift symmetry:
\begin{align}\label{eq:shift-symmetry}
    \phi(x)\mapsto \phi(x) + \alpha~, \qquad 0\leq \alpha < 2\pi~,
\end{align}
as well as a ``charge-conjugation'' symmetry $\phi\mapsto-\phi$. A potential typically breaks this symmetry, either completely or to a discrete subgroup. For instance, if $v(\phi)\propto \cos(2\phi)$, only the $\Z_2$ subgroup $\phi\mapsto\phi+\pi$ survives. 

Locally, $\phi$ is indistinguishable from its non-compact counterpart, the usual real massless scalar. Globally, however, the two theories differ. The relation \cref{eq:compactness} should be viewed as a gauge identification, and should not be conflated with the global $\U(1)$ symmetry \cref{eq:shift-symmetry}. As a result, $\phi$ is not a gauge-invariant operator in the quantum theory, but behaves like a scalar version of a gauge field. The well-defined gauge-invariant observables are built from derivatives, $\pd_\mu \phi(x)$ and higher derivatives, together with vertex operators,
\begin{align}
    V_n(x) = \ex{\ii n \phi(x)}~.
\end{align}
Note that the vertex operators are only gauge-invariant when $n\in\Z$.

Another defining property of the compact scalar theory is that, in contrast to the non-compact scalar, it allows for quantised winding. If spacetime contains a one-cycle, the scalar field can wind around its target-space circle. In differential form language, with $\dd{\phi} = \dd{x}^\mu \pd_\mu\phi$, the derivative of the compact scalar is quantised as
\begin{align}
    \int_C \dd{\phi} \in 2\pi\,\Z~,
\end{align}
for any oriented one-dimensional cycle, $C$, in spacetime. This is a variant of the Dirac, or flux, quantisation in electromagnetism. For global de~Sitter spacetime, this feature is relevant only for $\ds_2$. In higher dimensions it disappears, as $\ds_{D\geq 3}$ has no non-trivial one-cycles. However, this feature returns in other geometries with positive cosmological constant such as (extremal) Nariai black holes. 

With these definitions in place, we return to global $\ds_D$ and study the quantum theory of the free compact scalar \cref{eq:action-compact}, with $v=0$. We now summarise our main results.

\paragraph{Canonical quantisation: why UIRs are not enough.}
The UIRs of $\SO(D,1)$ describing massless minimally coupled scalars in $D$-dimensional de~Sitter spacetime are known; they were first identified in \cite{Higuchi:1986wu, Higuchi:1987hw}. They furnish the ``single-particle''%
\footnote{De~Sitter spacetime has no global notion of energy, but we nevertheless use the term particle for lack of a better word.\label{eq:particles?}}
Hilbert space, $\cH_{1}$, of the underlying QFT. But in making the leap from the single-particle Hilbert space to the full QFT, one may overlook sectors that are invisible to the UIRs. This is made clear in the work of Kirsten and Garriga \cite{Kirsten:1993ug}, who studied the non-compact scalar, and found that the full Hilbert space is not simply the Fock space, $\cF(\cH_1)$, over $\cH_1$. 

In this paper, we extend the Kirsten--Garriga quantisation to the case of the compact scalar, also known as an axion. We show in \cref{sec:canonical} that the axion Hilbert space is
\begin{align}\label{eq:H-axion}
    \cH_\t{axion} = L^2\qty\big(\S^1)\otimes \cF\qty(\cH_1)~,
\end{align}
where $L^2\qty\big(\S^1)$ is the quantum-mechanical Hilbert space of square-integrable functions on the circle. The states in $L^2\qty\big(\S^1)$ carry an integer angular momentum, $n\in\Z$, which is the conserved $\U(1)$ shift charge of the axion. We also determine the quantum $SO(D,1)$ charges and their action on the Hilbert space. Although \ds\ rotations retain the standard oscillator form, schematically $\sum_{n,m} \kappa_{nm}~ a^{\dagger}_n \, a_m$, the  dS boosts receive additional contributions from the quantum-mechanical sector, as in the non-compact case.

\paragraph{Zero-particle states and boosted observers.} In the Hilbert space \cref{eq:H-axion}, there is an infinite family of zero-particle
states, $\ket{\Omega_n} = \ket{n}_{\t{QM}} \otimes \ket{0}$. Only the $n=0$ state, $\ket{\Omega_0}$, is invariant under the whole $\SO(D,1)$ group. This is similar to the non-compact case where the only \ds-invariant zero-particle state is the one with vanishing continuous shift charge \cite{Kirsten:1993ug}. However, unlike in the non-compact case, all states $\ket{\Omega_n}$ are now normalisable.

We also observe an interesting phenomenon\footnote{This phenomenon has an analogue in the non-compact case at the mathematical level, but from a physical viewpoint,   acquires special interest in the compact case, as we explain in the paper.}.  In particular, although a geodesic observer measures zero quanta when the field is prepared in the state $\ket{\Omega_n}$, another geodesic observer related to the first one by a \ds\ boost will assign a non-zero particle number as long as $n \neq 0$. That is, dS transformations act on the states $\ket{\Omega_{n\neq 0}}$ by ``creating particles''.
{In particular, charged sectors do not admit a \ds-invariant notion of particle number, and as a consequence, different geodesic observers related through dS boosts assign different particle number to the same quantum state.} This is in sharp contrast with how dS transformations act on a \ds-invariant zero-particle state, such as the Bunch--Davies vacuum. We further compute field-strength Wightman functions in the states $\ket{\Omega_n}$, which we find to be Hadamard. For $n= 0$, they are  \ds-invariant. For $n\neq 0$, they are not \ds-invariant at finite times, but de~Sitter invariance is recovered asymptotically, at late (or early) times.

\paragraph{Euclidean path integrals.} We also ask how the Lorentzian charged sectors are captured by Euclidean sphere path integrals. In \cref{sec:sphere}, we first show how the ordinary $\S^D$ path integral $\parti_{\S^D}$ is expressed in terms of Harish-Chandra characters. However, the empty sphere path integral computes the neutral overlap $\braket{\Omega_0}$, and therefore captures only the oscillator Fock space. To capture individual charged sectors, as well as the entire Hilbert space, we compute sphere path integrals decorated appropriately by insertions of local operators. The decorated sphere path integral becomes a product of the empty sphere path integral times a quantum-mechanical partition function:
\begin{align}
    \widehat{\parti}_{\S^D} = \parti_\t{QM}\, \parti_{\S^D}~,
\end{align}
providing a Euclidean avatar of \cref{eq:H-axion}. Curiously, this path integral hints at a thermal behaviour with temperature $T_D$, which we identify as a function of $D$.

\paragraph{Duality and the $\ds_3$ photon.} The axion is dual to a $(D-2)$-form gauge potential. The corresponding \ds\ Hilbert spaces must therefore match. This raises a concrete question: how does the zero-mode sector of the scalar appear on the gauge-theory side? In \cref{sec:EM-Hilbert} we answer this question on $\ds_3$, where the compact scalar is dual to electromagnetism. In this case, the oscillator sector produces the photon UIR \cite{Loparco:2025azm}, while the axion's zero mode becomes a magnetic monopole sector of Maxwell theory. This sector has a topological origin, rooted in the fact that global $\ds_3$ has a non-trivial two-cycle, which can support magnetic flux.


\subsection{The many faces of axions}\label{ssec:axion-many-faces}

Before turning to the quantum theory of axions in de~Sitter, it is useful to spell out where such fields come from. This subsection contains well-known material for which we claim little to no originality, and serves mostly to contextualise our results to follow. The reader who is only interested in the details of the quantum theory of axions (compact scalars) in de~Sitter, may jump directly to \cref{sec:canonical}.

\paragraph{Goldstone bosons.}

A common origin is the phase of a complex order parameter. Take a complex scalar field, \(\Phi(x)\), charged under a global \(\U(1)\) symmetry, so that \(\Phi(x)\mapsto \ex{\ii \alpha}\Phi(x)\). We can parametrise it as
\begin{align}
    \Phi(x) = \rho(x)\,\ex{\ii \phi(x)}~.
\end{align}
The compact scalar is the angular part, $\phi$, of the complex scalar field. The original \(\U(1)\) symmetry then becomes the shift symmetry \cref{eq:shift-symmetry}, of $\phi$. If the radial mode is heavy, of order \(m_\rho\), for instance after spontaneous breaking of the \(\U(1)\) symmetry, then at scales well below \(m_\rho\) the effective action involves only \(\phi\). This is the usual massless Goldstone mode, with an action of the form \cref{eq:action-compact} and \(v=0\). In de~Sitter, this statement should be read as approximate: infrared effects tend to obstruct exact symmetry breaking or lead to pseudo-Goldstone modes \cite{Ratra:1985,Ford:1986,Boyanovsky:2012qs}. Nevertheless, pseudo-Goldstone axion-like angular fields are standard ingredients for inflationary effective theories \cite{Freese:1990rb,Arkani-Hamed:2003xts}.

\paragraph{Duals of higher-form fields.}

One can also arrive at a compact scalar by dualising a higher-form gauge field. A free massless compact scalar in $D$ dimensions is exactly dual to a $(D-2)$-form $\U(1)$ gauge field  \cite{Beasley:2014ila,Donnelly:2016mlc,Witten:2026twr} (cf. \cref{ssec:notation} for our differential form conventions). Let \(B\) denote the higher-form gauge field. We take its action to be (in Euclidean signature) 
\begin{align}\label{eq:B-action}
    S = \frac{1}{2h^2}\int \dd{B}\w\star \dd{B}~,
\end{align}
with quantised flux through every closed \((D-1)\)-dimensional surface \(\Sigma_{D-1}\):
\begin{align}
    \int_{\Sigma_{D-1}} \dd{B} \in 2\pi\,\Z~.
\end{align}
This theory admits a scalar description of the form \cref{eq:action-compact}, with \(v=0\), under the duality map%
\footnote{In Euclidean signature, the correct duality has a factor of $\ii$ as in \cref{eq:duality-D-dim}. If one naively substitutes \cref{eq:duality-D-dim} in \cref{eq:B-action} the dual action comes with the wrong kinetic term. The correct derivation of the duality involves introducing a Lagrange multiplier. Both of these remarks were reviewed recently in \cite{Witten:2026twr}. In Lorentzian signature there would be no factor of $\ii$.}
\begin{align}\label{eq:duality-D-dim}
    \dd{B} = \frac{\ii\, h^2}{2\pi} \star \dd{\phi}~,
\end{align}
with \(f=h/(2\pi)\). Non-perturbative effects, such as monopole proliferation, can then generate a periodic potential \cite{Polyakov:1976fu}. The compactness of \(\phi\) is essential here. If \(\phi\) were not compact, the dual theory would not be a standard \(\U(1)\) (higher-form) gauge field, but its infinite-coupling limit. Note that in this limit Wilson surface operators decouple.

In three dimensions, the duality identifies free Maxwell theory with a compact massless scalar. We will use this duality in \cref{sec:EM-Hilbert}, where we will spell out the topological subtleties in the Hilbert space of a photon in global \(\ds_3\).

\paragraph{Theta angle and the QCD axion.}

A paradigmatic source of compact scalars is quantum chromodynamics (QCD) in four dimensions, where the original axion was introduced as a possible solution to the strong CP problem \cite{Peccei:1977hh,Peccei:1977ur,Wilczek:1977pj,Weinberg:1977ma, Vacalis:2025uzk}\footnote{The axion is also a candidate for dark matter \cite{Preskill:1982cy, Abbott:1982af, Dine:1982ah}.}. The problem starts from the CP-violating term 
\begin{align}\label{eq:theta-term}
    S_\theta = \frac{\theta}{8\pi^2}\int \tr(G\w G)~,
\end{align}
which can be added to the QCD action, where $G = \frac{1}{2} G^{a}_{\mu\nu} T_a \dd{x}^\mu\w\dd{x}^\nu$ is the gluon field-strength two-form, with $T_a$ the $\su(3)$ generators. The coupling, $\theta$, is an angular parameter $\theta\sim\theta+2\pi$, because it multiplies the quantised instanton number.  The strong CP problem concerns the physical theta angle, $\bar{\theta} = \theta - \arg\det(Y_u Y_d)$, with $Y_{u,d}$ the up and down Yukawa matrices. In principle, $\bar{\theta}$ can take any value between $-\pi$ and $\pi$. Experiments instead bound it to be extremely small: $\abs*{\bar{\theta}} \lesssim 10^{-10}$ \cite{Harris:1999jx,Romalis:2000mg,Abel:2020gbr}. There is no principle within the minimal Standard Model that can explain this smallness. 

One possible solution introduces a compact scalar, $\phi(x)\sim\phi(x)+2\pi$,%
\footnote{In the literature this field is typically denoted by $a(x)$. For notational consistency with its other appearances, we denote it by $\phi(x)$.}
the ``axion''
\cite{Weinberg:1977ma,Wilczek:1977pj}. It is often viewed as a Goldstone mode of the ``Peccei--Quinn'' $\U(1)$ symmetry \cite{Peccei:1977hh,Peccei:1977ur}, acting by a chiral rotation on (some of) the quarks. On the axion itself, this symmetry appears as a shift symmetry, broken primarily by QCD instanton effects. The axion modifies the QCD action by 
\begin{align}
    \Delta S = S_\phi^\t{kin} + \frac{\kappa}{8\pi^2}\int \phi\,\tr(G\w G)~,
\end{align}
(see, e.g. \cite{Svrcek:2006yi}) where $S_\phi^\t{kin}$ denotes the kinetic term in \cref{eq:action-compact} and $\kappa$ is an integer. The axion shift symmetry \cref{eq:shift-symmetry} can then be used to absorb the theta term \cref{eq:theta-term} by $\phi\mapsto \phi - \theta/\kappa$. The effective CP violation is controlled by $\kappa\ev{\phi}$. The expectation value of the axion is computed after taking into account instanton effects, which generate a potential $v(\phi)$. In the standard QCD axion mechanism, this potential dynamically selects a CP-conserving point \cite{Peccei:1977hh,Peccei:1977ur,Vafa:1984xg,Svrcek:2006yi}, thus explaining the observed smallness of $\bar{\theta}$, i.e. absence of CP violation.

\paragraph{Dimensional reduction and string compactifications.}

A common mechanism for compact scalars comes from gauge fields wrapped on compact cycles \cite{Svrcek:2006yi} in higher dimensions. To illustrate, consider ordinary Kaluza--Klein (KK) reduction. Take a $(D+1)$-dimensional theory on $\S^1\times X_{D}$, containing a $\U(1)$ gauge field, $A$. After reduction to $X_D$, the holonomy of $A$ around the circle becomes a scalar field. More explicitly, if $x$ denotes coordinates on $X_D$ and $y$ is the coordinate on $\S^1$, the KK zero-mode is
\begin{align}
    \phi(x) = \int_{S^1} A(x,y)~.
\end{align}
Gauge invariance, $A \sim A +\dd{\chi}$, in particular under large gauge transformations, such that
\begin{align}
    \int_{\S^1} \dd{\chi} = 2\pi~,
\end{align}
implies that the scalar in the lower-dimensional theory is compact:
\begin{align}
    \phi(x) = \int_{S^1} A(x,y) \sim  \int_{S^1} A(x,y) + \int_{\S^1}\dd{\chi} = \phi(x) +2\pi~.
\end{align}

This mechanism extends to $p$-form gauge fields, integrated over compact $p$-cycles. This is the standard origin of axions in string compactifications. Such fields arise, for instance, from the Neveu--Schwarz (NS) 2-form, $B_2$, or from the Ramond--Ramond (RR) $p$-forms $C_p$, with $p=0,2,4$ in type IIB string theory and $p=1,3$ in type IIA, integrated over appropriate cycles, as well as model-dependent heterotic string axions \cite{Svrcek:2006yi}. This mechanism gives rise to a raft of axions, determined by the topology of the compactification manifold. For instance, from the NS 2-form alone, one gets as many axions as there are 2-cycles. For example, on a six-torus, this gives $15=\binom{6}{2}$ compact scalars from this sector alone. 
Furthermore, this mechanism predicts that the axion potential vanishes to any order in perturbation theory.%
\footnote{At the level of the higher-dimensional supergravity action, this protection is usually traced to gauge invariance \cite{Svrcek:2006yi}. It would be conceptually cleaner to explain its origin using a global symmetry. However, in supergravity continuous electric higher-form symmetries \cite{Gaiotto:2008ak} are typically broken by Chern--Simons couplings. These symmetries can be resurrected as non-invertible symmetries \cite{Garcia-Valdecasas:2023mis}. It is likely that the perturbative shift symmetry of the four-dimensional axion is protected by non-invertible symmetries of the parent supergravity.}
Non-perturbative effects can still generate a potential. See, for instance \cite{Svrcek:2006yi,Blumenhagen:2009qh,Kachru:2003aw,Wen:1985jz,Guidetti:2022xct}. 

If one wants the lower-dimensional spacetime to be de~Sitter, one faces a further question apart from the origin of the axions, namely the origin of de~Sitter itself. There are candidate de~Sitter compactifications, for instance \cite{Kachru:2003aw,Balasubramanian:2005zx}, but their status is still debated.  It is also worth mentioning that extra-dimensional origins of axions are of interest in cosmology, appearing in certain models of inflation \cite{Arkani-Hamed:2003xts}. 

\paragraph{Discrete gauging of non-compact fields.}

Another possibility is that the scalar started its life as a non-compact field, for instance as a Goldstone mode of a non-compact global symmetry. Effective theories of such Goldstone modes are often problematic and can appear non-local \cite{Anninos:2021ydw,Anninos:2023lin}. One way out is to keep the non-compact symmetry, but embed the Goldstone EFT into a larger gauge theory and study non-local observables \cite{Anninos:2021ydw}. Another is to gauge the entire non-compact shift symmetry of the Goldstone mode \cite{Anninos:2023lin}. Here we give an intermediate option: gauge only a discrete subgroup of the shift symmetry.

To be explicit, consider a real non-compact scalar, $\Phi$, with Euclidean action 
\begin{align}
    S[\Phi] = \frac{1}{2}\int \dd[D]{x}\, \sqrt{\abs{g}}\; g^{\mu\nu} \pd_\mu \Phi \, \pd_\nu \Phi = \frac{1}{2}\int \dd{\Phi}\w\star\dd{\Phi}~.
\end{align}
This action has a global non-compact shift symmetry, with group $\R$, where $\Phi\mapsto\Phi+c$, for any $c\in \R$. We may gauge a subgroup $\Z\subset \R$, for instance, shifts by integer multiples of $h=2\pi f$, to choose a convenient normalisation. Gauging this subgroup requires a discrete gauge field. It is more convenient to adapt the method of \cite{Argurio:2024ewp} to $D>2$ by introducing instead a non-compact, $\R$-valued, 1-form gauge field $a$. This gauge field is then restricted to a $\Z$ gauge field by a Lagrange multiplier, namely a $\U(1)$ $(D-2)$-form gauge field $B$.%
\footnote{This is superficially similar to \cite{Anninos:2023lin}, with a small but all-important difference, namely that $B$ is a \emph{compact} $(D-2)$-form gauge field, while \cite{Anninos:2023lin} introduce an $\R$-valued field.}
The gauged action reads:
\begin{align}
    S[\Phi,a,B] = \frac{1}{2}\int \qty(\dd{\Phi}-a)\w\star\qty(\dd{\Phi}-a) - \frac{\ii}{h}\,\int \dd{B}\w a~.
\end{align}
Recall that, here, the 0-form $\Phi$ is non-compact, the $(D-2)$-form gauge field is compact (in the sense that $\int B\sim\int B+2\pi$), while $a$ is a non-compact 1-form.
The second term in the action enforces that $a$ has no curvature and $\int a \in h\,\Z$, making it a $\Z$ gauge field. Integrating out $a$ gives
\begin{align}\label{eq:Dual-theory}
    S[B] = \frac{1}{2h^2} \int \dd{B}\w\star\dd{B}~, \qq{with} \int\dd{B}\in2\pi\,\Z~,
\end{align}
see \cite{Argurio:2024ewp} for further details. Using the duality between a compact scalar and a $(D-2)$ form, we recognise \cref{eq:Dual-theory} as a dual presentation of an axion with decay constant $f = h/(2\pi)$.

\paragraph{Bosonisation.}

In two dimensions, fermionic theories can often be written in bosonic variables. Formally, this is done by gauging the fermion number symmetry $(-1)^F$ in the fermionic model, a procedure known as ``bosonisation'' \cite{Coleman:1974bu,Mandelstam:1975hb,Karch:2019lnn}. The paradigmatic example is a free massless Dirac fermion, which maps to a free compact minimally coupled massless scalar, at a specific value of the coupling, $f$. Deformations away from this duality give interacting models built out of the compact scalar. For instance, the massive Thirring model bosonises to the sine-Gordon model, where the compact scalar acquires a potential $v(\phi)\sim \cos\phi-1$. For de~Sitter, bosonisation dualities have recently served as a source of inspiration for exactly solvable models \cite{Anninos:2024fty,Aguilera-Damia:2026dbk,Galati_Vitouladitis}.

\paragraph{Lattice models.}

For completeness, we finally mention that compact bosons appear frequently in the infrared of lattice models, such as quantum rotor  and XY models \cite{Kosterlitz:1973xp,Savit:1979ny}. In these examples, the microscopic variable is a phase at each lattice site, so compactness is built into the model itself. Finally, the Villain and modified Villain formulation \cite{Villain:1974ir,Sulejmanpasic:2019ytl,Gorantla:2021svj} of such lattice models connects directly with dualities, see e.g. \cite{Argurio:2026txf,Bacq:2026ega}.

\subsection{Notation and conventions}\label{ssec:notation}

For the reader's convenience we collect here the notation used frequently throughout the paper.

\paragraph{Units.}

We denote the de~Sitter radius by \(\rds\), with units of length. We take the axion to be dimensionless $\qty[\phi] = \rds^{0}$. The dimensions are carried by the axion decay constant $f^2$ as \(\qty[f^2]=\rds^{2-D}\).

\paragraph{Global de~Sitter.} The line element for global $\ds_D$ is \begin{align}\label{eq:global-dsd-metric}
    \dd{s}^2 = \rds^2\qty(-\dd{t}^2+\cosh^2t\,\dd{\Omega}_{D-1}^2)~,
\end{align}
where \(t\) is dimensionless and $\dd \Omega_{D-1}^2$ is the line element of the unit $S^{D-1}$. The line element of $S^{j}$, for $j=D-1,D-2,\ldots,2$, is expressed recursively as
\begin{alignat}{3}
    \dd{\Omega}^2_{j} &= \dd{\theta}_{j}^2+\sin^2{\theta_{j}}\,\dd{\Omega}_{j-1}^2,\qquad && 0\leq\theta_{j} \leq \pi~, \\
    \dd{\Omega}^2_{2} &= \dd{\theta}_{2}^2+\sin^2{\theta_{1}}\,\dd{\theta_1}^2, \qquad && 0\leq \theta_1 \leq 2\pi~.
\end{alignat}
The volume element of the unit $S^{D-1}$ is
\begin{equation}
    \dvol_{D-1}=\sqrt{\tilde{g}}\,\dd[D-1]{\theta} \equiv \sqrt{\tilde{g}}\,\dd{\theta}_{D-1}\cdots\dd{\theta_{2} }\dd{\theta_1}~,
\end{equation}
with total volume
\begin{align}\label{define vol_(D-1) dimensionless}
    \Vol_{D-1} \coloneqq \int_{\S^{D-1}}\dvol_{D-1}
    = \frac{2\pi^{D/2}}{\Gamma(D/2)}~.
\end{align}
Our convention for the totally antisymmetric tensor, \(\epsilon_{\mu_{1}\mu_{2}\ldots\mu_{D}}\), namely the components of the volume form on $\ds_D$ is 
\begin{align*}
    \epsilon_{t \theta_1 \theta_2\ldots\theta_{D-1}} = \sqrt{-g}~, \quad\qq{and}\quad \epsilon_{t j_1\ldots j_{D-1}}=\cosh^{D-1}{t}\; \tilde{\epsilon}_{j_1 \ldots j_{D-1}}~,  
\end{align*}
where $\tilde{\epsilon}_{j_1 \ldots j_{D-1}}$ is the antisymmetric tensor on the unit $S^{D-1}$. Lowercase Greek indices denote components in the global $\ds_D$ coordinates \cref{eq:global-dsd-metric}, $\mu,\nu,\ldots\kern-2pt \in\kern-2pt \{t, \theta_{D-1},\ldots,\theta_2, \theta_1   \}$. Lowercase Latin indices denote components along the spatial $S^{D-1}$: $i,j,k,\ldots \in \{ \theta_{D-1},\ldots,\theta_2, \theta_1\}$. The square root of the determinant of the de~Sitter metric is 
\begin{align*} 
    \sqrt{-g} =  \cosh^{D-1}{t}\;\sqrt{\tilde{g}}~,
\end{align*}
where $\tilde{g}$ is the determinant of the metric on the unit $S^{D-1}$. Note that $\tilde{\epsilon}_{\theta_1  \theta_2\ldots\theta_{D-1} } = \displaystyle\sqrt{\tilde{g}}$. 
We use \(\vec{\theta}\) as a shorthand for the angular coordinates \((\theta_{D-1},\ldots,\theta_2,\theta_1)\) on $S^{D-1}$. 

Covariant derivatives, $\nabla_\mu$, on $\ds_D$ and $\tilde{\nabla}_i$ on the unit $\S^{D-1}$ have dimensions $[\nabla_{\mu}] = \rds^{-1}$ and $[\tilde{\nabla}_{i}] = \rds^0$, respectively. 
We also  use the notation 
\begin{align}
    \dd[D]{x} = \rds^D \dd{t}\,\dd\theta_{D-1}\cdots \dd{\theta_2}\,\dd{\theta_1}~.    
\end{align}
The global coordinates on $\ds_D$  are taken to be dimensionless. We define $x^t = \rds \,t$, so 
\begin{equation}
    \pd_t = - \partial^t = \rds^{-1}~\pdv{t}~,
\end{equation}
and similarly $x^{\theta_{D-1}} = \rds \, \theta_{D-1},\ \ldots,\ x^{\theta_1} = \rds \, \theta_1$.

\paragraph{Killing vectors.} The isometry algebra of $\ds_D$ is $\so(D,1)$, generated by $(D+1)D/2$ Killing vectors. The $D(D-1)/2$ rotational Killing vectors, $R^{\mu}$, generate the maximally compact subalgebra $\so(D)$ and coincide with the Killing vectors of the spatial $S
^{D-1}$. The $D$ remaining Killing vectors, $B^{\mu}$, generate boosts. A convenient basis for the boosts is written in terms of the $\ell=1$ scalar spherical harmonics $Y_{\ell \vec{m}}(\vec{\theta})$ on the unit $S^{D-1}$ defined in eq. \cref{eq:spherical-harmonics-dsd} (see, e.g., \cite{Gibbons-Higuchi})
\begin{equation}\label{def: boost Killing vectors}
\begin{split}
    & B_{\vec{m}}^{t} (t,\vec{\theta})= B_{\vec{m}}^{t} (\vec{\theta})=Y_{\ell=1,\vec{m}}(\vec{\theta})  \\
    & B_{\vec{m}}^{j} (t,\vec{\theta})=\tanh{t}~\tilde{\nabla}^{j}Y_{\ell=1,\vec{m}}(\vec{\theta})~,
\end{split}
\end{equation}
up to an overall normalisation constant. The label $\vec{m}$ is defined below eq. \cref{eq:spherical-harmonics-dsd}.

Unless stated otherwise, $\xi^{\mu}$ denotes a general dS Killing vector, while $B^{\mu}$ and $R^{\mu}$ denote boost and rotation Killing vectors, respectively. We take Lie derivatives  to be dimensionless. For example, for a scalar field $\phi$,
\begin{align}
    \lie_{\xi} \phi \coloneqq \rds~\xi^{\nu} \frac{\partial}{\partial x^{\nu}}~.
\end{align}

\paragraph{Euclidean sphere.} We use polar coordinates on the Euclidean sphere \(\S^D\),
\begin{align}\label{eq:euclidean-sphere-metric}
    \dd{s}_\t{E}^2
    =
    \rds^2\qty(\dd{\tau}^2+\sin^2\tau\,\dd{\Omega}_{D-1}^2)~,
    \qquad
    0\leq\tau\leq\pi~.
\end{align}
The angular coordinates \(\vec{\theta}\) are the same coordinates used above on the spatial \(\S^{D-1}\). We denote the north and south poles by
\begin{align}
    \t{N}:\,\tau=0~,
    \qquad
    \t{S}:\,\tau=\pi~.
\end{align}
The line element of global $\ds_D$ is related to that of $S^D$ by analytic continuation as
\begin{align}\label{eq:euclidean-lorentzian-continuation}
    \tau=\frac{\pi}{2}-\ii t~,
    \qquad
    \sin\tau=\cosh t~.
\end{align}
The volume element of $S^D$ is
\begin{align}
    \sqrt{g_\t{E}}\,\dd[D]{x_\t{E}}
    =
    \rds^D\sin^{D-1}\tau\,\dd{\tau}\,\dvol_{D-1}~.
\end{align}
For two points \(x=(\tau_x,\vec{\theta}_x)\) and \(y=(\tau_y,\vec{\theta}_y)\) on $S^D$, we use the invariant distance
\begin{align}\label{eq:sphere-invariant-xi}
    \xi(x,y)
    =
    \cos(\frac{\ell_{\S^D}(x,y)}{\rds})
    =
    \cos\tau_x\cos\tau_y
    +
    \sin\tau_x\sin\tau_y\,
    \cos{\hat{\ell}}_{\S^{D-1}}(\vec{\theta}_x,\vec{\theta}_y)~.
\end{align}
Here $\ell_{\S^D}$ is the geodesic distance on $\S^D$ with radius $\rds$ and \(\hat{\ell}_{\S^{D-1}}\) is the geodesic distance on the unit \(\S^{D-1}\). Antipodal points have \(\xi=-1\).

\paragraph{Differential forms.} Conventionally, we normalise differential \(p\)-forms with a factor of \(1/p!\) when spelt out in components:
\begin{align*}
    \omega = \frac{1}{p!}\omega_{\mu_1\mu_2\cdots\mu_{p}}(x)\dd{x^{\mu_1}}\w\dd{x^{\mu_2}}\w\cdots\w\dd{x^{\mu_p}}~.
\end{align*}
Differential forms are always dimensionless. Dimensions are carried by their components and the Hodge-star operator, $\star$.

\paragraph{Conventions for $D=3$.} For $D=3$, we use the convention
\begin{align*}
    \sin{\theta_2} \dd{\theta_2} \w \dd{\theta_1} = -\frac{1}{2}  \frac{\tilde{\epsilon}_{jk}}{\rds^2} \dd{x^{j}} \w \dd{x^{k}}~,
\end{align*}
for the volume form of the spatial $\S^2$, with $\int_{S^2} \sin{\theta_2}\, \dd{\theta_2} \w \dd{\theta_1} = 4\pi$. The fields relevant for three-dimensional electromagnetism have dimensions $\qty[A_{\mu}]= \rds^{-1}$ and $\qty[F_{\mu \nu}] = \rds^{-2}$.


\section{The compact scalar Hilbert space on \tps{\ds_D}{dSD}}\label{sec:canonical}

In this section, we quantise the massless minimally coupled compact scalar field on global $\ds_D$ following closely the approach of Kirsten and Garriga \cite{Kirsten:1993ug} for the case of the non-compact massless scalar. See also \cite{Gibbons-Higuchi-Yang-Mills, Tolley:2001gg, Allen:1985ux, Allen:1987tz, Letsios:2024snc}.%
\footnote{For a different method of quantisation (related to the Gupta--Bleuler quantisation of electromagnetism) for the massless minimally coupled scalar field, as well  as for ``tachyonic'' scalars, in de~Sitter spacetime see \cite{Epstein:2014jaa, Bertola:2006df}. See also \cite{Anninos:2023lin} for the quantisation of a massless scalar field with gauged shift symmetry in two dimensions.}
Unlike the non-compact case, the \ds-invariant vacuum state  for the compact scalar is normalisable.
In view of the aforementioned duality between electromagnetism and the compact massless scalar field on $\ds_3$, the Hilbert space we construct (which is not a standard Fock space) for the compact massless scalar also defines the Hilbert space of the free quantum electromagnetic field on $\ds_3$. Similar comments apply to the case of $\ds_D$ ($D \geq 4$) where the compact massless scalar theory is dual to a theory of a $(D-2)$-form gauge potential.

We consider the free compact scalar $\phi\sim\phi+2\pi$, with Lorentzian action
\begin{align}\label{eq:compact-scalar-action-dsd}
    S[\phi]
    = -\frac{f^2}{2}\int\dd[D]{x}\sqrt{-g}\;g^{\mu\nu}\pd_\mu\phi\,\pd_\nu\phi~.
\end{align}
The axion decay constant, $f^2$, has units of  \(\rds^{2-D}\). It is useful to define the dimensionless coupling constant
\begin{align}
    \kappa_D \coloneqq \rds^{D-2}f^2~.
\end{align}
We will work in global coordinates \cref{eq:global-dsd-metric}.
In these coordinates, the action \cref{eq:compact-scalar-action-dsd} becomes
\begin{align}\label{eq:compact-scalar-action-global}
    S[\phi]
    =
    \frac{\kappa_D}{2}\int\dd{t}\,\dd[D-1]{\theta}~\sqrt{\tilde{g}}~\,\cosh^{D-1}t\,
    \qty[
        \qty(\pdv{\phi}{t})^2
        -\frac{1}{\cosh^2t}\,\tilde{g}^{\,ij}\,\tilde{\nabla}_i\phi\; \tilde{\nabla}_j\phi
    ]~.
\end{align}


\subsection{Classical modes}

The field equation that follows from \cref{eq:compact-scalar-action-dsd} is
\begin{align}\label{field equation massless scalar}
    \Box\,\phi=0~.
\end{align}
We are interested in the set of mode solutions of this equation that can be obtained by analytically continuing the spherical harmonics on $S^D$, as described in, e.g., \cite{Higuchi:1986wu}. These modes are often called  ``Bunch--Davies'' modes because of their use in the definition of the Bunch--Davies vacuum (discovered earlier in \cite{AIHPA_1976__25_1_67_0, Chernikov:1968zm}).

In global coordinates, eq. \cref{field equation massless scalar} reads \cite{Higuchi:1986wu}:
\begin{align}\label{eq:kg-dsd-global}
    \qty(
        -\frac{\pd^2}{\pd t^2}
        -(D-1)\tanh t\,\frac{\pd}{\pd t}
        +\frac{\tilde{\Box}}{\cosh^2t}
    )\phi(t,\vec{\theta})=0~,
\end{align}
where \(\tilde{\Box}\) is the Laplacian on the unit \(\S^{D-1}\). Let us introduce the scalar spherical harmonics \(Y_{\ell\vec{m}}(\vec{\theta})\). These obey \cite{Higuchi:1986wu}
\begin{align}\label{eq:spherical-harmonics-dsd}
    \tilde{\Box}\, Y_{\ell\vec{m}}
    = -\ell(\ell+D-2)Y_{\ell\vec{m}}~,
\end{align}
and are orthonormal:
\begin{align}
    \int_{\S^{D-1}}\dvol_{D-1}\,
    Y_{\ell\vec{m}}(\vec{\theta})^*~Y_{\ell''\vec{m}''}(\vec{\theta})
    = \delta_{\ell\ell''}\delta_{\vec{m}\vec{m}''}~.
\end{align}
In the above, $\ell = 0,1,2,\ldots$ and $\tilde{\Box} = \tilde{g}^{\,ij}\, \tilde{\nabla}_i \tilde{\nabla}_j$ is the Laplace-Beltrami operator on $S^{D-1}$.
The multi-index \(\vec{m}\) labels the degeneracy of scalar harmonics at fixed \(\ell\). In other words, it keeps track of the dimension of the $\SO(D)$ representation with highest weight $(\ell,0,\ldots,0)$. We have
\begin{align}\label{eq:labels-spherical-harmonics}
    \vec{m}=(m_{D-2},\ldots,m_{1})~,
    \qquad
    \ell\geq m_{D-2}\geq\cdots\geq m_{2}\geq |m_{1}|~.
\end{align}
Here the entries before the last one are non-negative integers and \(m_{1}\in\Z\). For \(D=3\), the chain reduces to the usual condition \(-\ell\leq m_1\leq\ell\). The number of allowed \(\vec{m}\)'s, i.e. the dimension of the corresponding $\SO(D)$ representation with highest weight $(\ell,0,\ldots,0)$, is \cite{Camporesi:1990wm}:      
\begin{align}
    N_{\ell,D}
    =
    \frac{(2\ell+D-2)(\ell+D-3)!}{\ell!(D-2)!}~,
\end{align}
with \(N_{0,D}=1\),  \(N_{1,D}=D\). The constant harmonic is \(Y_{0 \vec{0}}=\Vol_{D-1}^{-1/2}\).

\paragraph{$\ell \geq 1$ sector.}

For \(\ell\geq 1\), the Bunch--Davies modes  obtained by analytic continuation from $S^{D}$ are \cite{Chernikov:1968zm,Allen:1985ux,Bunch:1978yq,Marolf:2008hg, Higuchi:1986wu}
\begin{align}\label{eq:bd-modes-dsd}
    \phi_{\ell\vec{m}}(t,\vec{\theta})
    =
    \cN_{\ell,D}\,
    \qty(\cosh t)^{-\frac{D-2}{2}}\,
    P_{\frac{D-2}{2}}^{-\qty(\ell+\frac{D-2}{2})}(\ii\sinh t)\,
    Y_{\ell\vec{m}}(\vec{\theta})~,
\end{align}
where \(P_\nu^\mu(z)\) is the associated Legendre function of the first kind. The normalisation factor is
\begin{align}
    \cN_{\ell,D}
    =
    \qty(\frac{\Gamma(\ell)\Gamma(\ell+D-1)}{2})^{1/2}~,\qquad \ell\geq 1~.
\end{align}
The modes are normalised by the Klein--Gordon inner product, defined as
\begin{align}\label{KG scalar product}
    \kip{\phi^{(1)}}{\phi^{(2)}}
    =~
    \ii\cosh^{D-1}t\int_{\S^{D-1}}\dvol_{D-1}\,
    \qty[
        \qty(\phi^{(1)})^*~\frac{\pd}{\pd t}\phi^{(2)}
        -\qty(\frac{\pd}{\pd t}\phi^{(1)})^*\,\phi^{(2)}
    ]~.
\end{align}
On solutions of \cref{eq:kg-dsd-global}, this scalar product is both time-independent and \ds-invariant~\cite{Higuchi:1986wu}. In particular, in terms of the modes \(\phi_{\ell\vec{m}}\) we have:
\begin{align}
    \kip{\phi_{\ell\vec{m}}}{\phi_{\ell''\vec{m}''}}
    = \delta_{\ell\ell''}\delta_{\vec{m}\vec{m}''}~,
    \quad
    \kip{\phi_{\ell\vec{m}}^*}{\phi_{\ell''\vec{m}''}^*}
    = -\delta_{\ell\ell''}\delta_{\vec{m}\vec{m}''}~,
    \quad
    \kip{\phi_{\ell\vec{m}}}{\phi_{\ell''\vec{m}''}^*}=0~.
\end{align}
The (infinitesimal) dS invariance of the scalar product means that Lie derivatives with respect to Killing vectors, $\lie_{\xi}$, are anti-hermitian, as~\cite{Higuchi:1986wu}
\begin{align}\label{dS-d invariance of KG scalar product}
    \kip{\lie_{\xi} \phi^{(1)}}{ \phi^{(2)}} +  \kip{\phi^{(1)}}{ \lie_{\xi}\phi^{(2)}} = 0.
    \end{align}
The Klein--Gordon scalar product can be understood as the time-independent Noether charge associated with the conserved Klein--Gordon current:
\begin{align}\label{def: KG current dS-d}
    J^{\mu}_\t{KG}\qty(\phi^{(1)},\phi^{(2)}) &= -i\,\rds\,\qty[\qty(   \phi^{(1)})^{*}\, \nabla^{\mu}  \phi^{(2)} - \qty( \nabla^{\mu}\,\phi^{(1)})^{*}\,   \phi^{(2)}]~, \\ 
    \nabla_{\mu}J^{\mu}_\t{KG}\qty(\phi^{(1)},\phi^{(2)}) &=0~,
\end{align}
that is:
\begin{align}
    \kip{\phi^{(1)}}{\phi^{(2)}}    = \int_{S^{D-1}} \dd[D-1]{\theta}\, \sqrt{-g}\,  J^{t}_\t{KG}\qty(\phi^{(1)},\phi^{(2)})~.
   \end{align}

\paragraph{Positive frequency condition.} The Bunch--Davies modes, $\phi_{\ell \vec{m}}(t, \vec{\theta})$, as in \cref{eq:bd-modes-dsd}, can be viewed as the analogues of positive frequency modes. In particular, in the short wavelength limit ($\ell \gg 1$), they behave as flat-space positive frequency modes%
\footnote{Similarly, their complex conjugates, $\phi_{\ell m}(t, \vec{\theta})^{*}$, are the analogues of negative frequency modes. Positive frequency Bunch--Davies modes do not mix with negative frequency modes under dS transformations. This will be demonstrated shortly.}
\begin{align}\label{positive freq condition ell>>1}
    \frac{\partial}{ \partial t} \phi_{\ell \vec{m}}(t, \vec{\theta}) \sim - i \frac{\ell}{\cosh{t}} \phi_{\ell \vec{m}}(t, \vec{\theta})~, \qquad \ell \gg 1.
\end{align}
 This short wavelength behaviour of modes is standard in QFT in global dS spacetime. For example, in the case of the graviton \cite{Higuchi:1991tn, Higuchi:2002sc} and spinor fields \cite{Letsios:2020twa}, it ensures that the Wightman two-point function, constructed through the mode-sum method, satisfies the Hadamard condition \cite{Fewster:2013lqa}.


\paragraph{Zero-mode sector.} The zero-mode sector must be treated separately. In particular, the equation for the mode with $\ell=m_{D   -2}=\cdots=m_2 = m_1=0$ is
\begin{align}
    \frac{\pd}{\pd t}\qty(\cosh^{D-1}t\,~\frac{\pd}{\pd t}\phi_{0\vec{0}})=0~.
\end{align}
There are two independent zero-mode solutions. One is a constant, $1$, 
and the other a time-dependent solution, \(f_D(t)\), where
\begin{align}\label{time-dep zero-mode dS-d}
    f_D(t) \coloneqq \frac{1}{\Vol_{D-1}}\int_0^t\frac{\dd{s}}{\cosh^{D-1}s}~,
    \qquad
    \frac{\pd}{\pd t} f_D(t)=\frac{1}{\text{vol}_{D-1}}\cosh^{-(D-1)}t .
\end{align}
Equivalently, this can be written in closed form as\footnote{For integer \(D\), one may also use the recursion
\begin{align}
   \Vol_{D-1} f_D(t)
    =
    \frac{\tanh t\,(\cosh t)^{-(D-3)}}{D-2}
    +
    \frac{D-3}{D-2}~\Vol_{D-3}~f_{D-2}(t)~,
\end{align}
with \(f_2(t)= \frac{1}{\Vol_1}\arctan(\sinh t)=\frac{1}{\Vol_1}\arcsin(\tanh t)\) and \(f_3(t)=\frac{1}{\Vol_2}\tanh t\). For example,
\begin{align}
    f_4(t)
    =\frac{1}{\Vol_3}
    \frac{1}{2}\qty(\tanh t\,\cosh^{-1}t+\arctan(\sinh t))~,
    \qquad
    f_5(t)
    = \frac{1}{\Vol_4}
    \tanh t-\frac{1}{3}\tanh^3t~.
\end{align}}
\begin{align}
    f_D(t)
    =\frac{1}{\Vol_{D-1}}
    \tanh t\;
    {}_2F_1\qty(
        \frac{1}{2},
        \frac{3-D}{2};
        \frac{3}{2};
        \tanh^2 t
    )~.
\end{align}
For \(D=3\), this reduces to \(f_3(t)=\frac{1}{\Vol_2}\tanh t\). The inner product for the zero-mode solutions is
\begin{align}
    \kip{1}{1}=0~,
    \qquad
    \kip{f_D}{f_D}=0~,
    \qquad
    \kip{f_D}{1}=-\ii\,~.
\end{align}
Note that the zero-mode sector is orthogonal to the $\ell\geq 1$ sectors:
\begin{align}
    \kip{\phi_{\ell\vec{m}}}{1} = \kip{\phi_{\ell\vec{m}}}{f_D}=\kip{\phi_{\ell\vec{m}}^*}{1} = \kip{\phi_{\ell\vec{m}}^*}{f_D}
    =0~.
\end{align}

\subsection{Canonical quantisation}

The momentum density conjugate to \(\phi\), determined from the action functional, is
\begin{align}\label{eq:canonical-momentum-dsd}
    \Pi(t,\vec{\theta})
    =
    \kappa_D~\sqrt{\tilde{g}}~\cosh^{D-1}t\,~\frac{\pd}{\pd t}\phi(t,\vec{\theta})~.
\end{align}
We decompose the quantum field $\hat{\phi}$ into its zero-mode $\hat{\phi}_{0}$ and oscillator $\hat{\varphi}$ parts, as
\begin{align}
    \hat{\phi}(t,\vec{\theta}) = \hat{\phi}_{0}(t,\vec{\theta}) +\hat{\varphi}(t,\vec{\theta}).
\end{align}
At equal times, the canonical commutation relation is
\begin{align}
    \comm{\hat{\phi}(t,\vec{\theta})}{\hat{\Pi}(t,\vec{\theta}')}
    =
    \ii\,\delta(\vec{\theta},\vec{\theta}')~,
\end{align}
where $\delta(\vec{\theta},\vec{\theta}')$ is defined by \(\int_{\S^{D-1}} \dd[D-1]{\theta'}\,\delta(\vec{\theta},\vec{\theta}')f(\vec{\theta}')=f(\vec{\theta})\). Note that the integration measure is simply $ \dd[D-1]{\theta'}$ without an explicit factor of $\sqrt{-g}$.

The spatially constant part of \(\hat{\Pi}\) is the conserved shift charge,
\begin{align}\label{eq:shift-charge-dsd}
    \hat{p}
    =
    \int_{\S^{D-1}} d^{D-1}\theta\,~\hat{\Pi}(t, \vec{\theta})
    =
    \kappa_D \Vol_{D-1}\cosh^{D-1}t\,~\frac{\pd}{\pd t}\hat{\phi}_0(t)~.
\end{align}
It can be readily shown that $\partial \hat{p} / \partial t=0$, as a direct consequence of the field equations.

The zero-mode field operator is
\begin{align}\label{quantised zero mode}
    \hat{\phi}_0(t)
    =
    \hat{q}
    +
    \frac{\hat{p}}{\kappa_D}\,f_D(t)~,
    \qquad
    \comm{\hat{q}}{\hat{p}}=\ii~,
    \qquad
    \hat{q}\sim \hat{q}+2\pi~.
\end{align}
The compactness of \(\hat{\phi}\) means that \(\hat{q}\) is an angle. Therefore, the zero-mode operator acts on the quantum-mechanical Hilbert space of the particle on the circle. As a result, the ``angular momentum'' operator \(\hat{p}\) has integer spectrum:
\begin{align}\label{eigenstates of p}
    \hat{p}\ket{n}_{\t{QM}}=n\ket{n}_{\t{QM}}~,
    \qquad
    n\in\Z~,
\end{align}
where the subscript $\t{QM}$  stands for quantum-mechanical. A useful representation is\footnote{See, e.g., \cite{Anastopoulos_2023}.}
\begin{alignat}{2}
   \hat{p} &=-\ii\pd_q~,
    \quad
    \psi_n(q)=\,_\t{QM}\!\braket{q}{n}_\t{QM}=\frac{1}{\sqrt{2\pi}}\ex{\ii nq}~,
    \quad 
    \int_{0}^{2\pi}\dd{q} \psi^*_{n}(q) \psi_{n'}(q)= \delta_{nn'}.
\end{alignat}
These states span the Hilbert space of square-integrable functions on a circle, $L^2\qty\big(S^1)$.

Strictly speaking, \(\hat{q}\) itself is a coordinate on a patch of the circle. The globally defined operators are $\ex{\ii n \hat{q}}$. These are naturally acted upon by  exponentials  $\hat U_\alpha = \ex{\ii \alpha \hat{p}}$, as
\begin{align}\label{eq:clock}
   \hat  U_\alpha ~\ex{\ii n \hat{q}}~ \hat U_\alpha^{-1}=\ex{\ii\alpha n}\,\ex{\ii n \hat{q}}~,
\end{align}
with \(n\in\Z\) and \(\alpha\sim\alpha+2\pi\) (equivalently \(\alpha\in\U(1)\)). In particular,
\begin{align}
    \ex{\ii n \hat{q}}\ket{n'}_{\t{QM}}=\ket{n+n'}_{\t{QM}}~.
\end{align}
Tracing back the origin of the \(\ex{\ii m \hat{q}}\) operators, we see that they are exactly the zero-mode contribution of the vertex operators of the compact scalar field \( \hat{V}_{m}=\no{\ex{\ii m \hat \phi}}\). 
The zero-mode Hamiltonian associated with translations of the global time coordinate is time-dependent, 
\begin{align}\label{eq:particle-ring-hamiltonian}
    \hat{H}_0(t)
    =
    \frac{\hat{p}^2}{2\,I\, \cosh^{D-1}t}~, \qq{where} I = \rds\,\kappa_D\,\Vol_{D-1}~.
\end{align}
This is the Hamiltonian of a simple quantum rotor with time-dependent moment of inertia, $I(t) = I\, \cosh^{D-1}t$. This time dependence is not a problem, as global time translations are not an isometry of de~Sitter.

The oscillator part of the field is expanded in $\ell \geq 1$ modes as
\begin{align}\label{quantised oscillator part}
    \hat{\varphi}(t,\vec{\theta})
    =
    \frac{1}{\sqrt{\kappa_D}}
    \sum_{\ell=1}^{\infty}\sum_{\vec{m}}
    \qty(
        \hat a_{\ell\vec{m}}\phi_{\ell\vec{m}}(t,\vec{\theta})
        +
        \hat a_{\ell\vec{m}}^{\dagger}\phi_{\ell\vec{m}}^*(t,\vec{\theta})
    )~.
\end{align}
It follows that the coefficients $\hat{a}_{\ell\vec{m}}$ and $\hat{a}^\dagger_{\ell\vec{m}}$ obey the standard commutation relations:
\begin{align}
    \comm{\hat a_{\ell\vec{m}}}{\hat a_{\ell''\vec{m}''}^\dagger}
    = \delta_{\ell\ell''}\delta_{\vec{m}\vec{m}''}~.
\end{align}
Annihilation operators can also be expressed as:
\begin{align}\label{annihilation operator in terms of KG inn prod ds-d}
    \hat a_{\ell\vec{m}}
    =
    \sqrt{\kappa_D}\,\kip{\phi_{\ell\vec{m}}}{\hat\varphi}~,
\end{align}
and similarly for $\hat{a}^\dagger_{\ell\vec{m}}$. The full quantum field is
\begin{align}\label{eq:full-compact-scalar-mode-expansion}
    \hat{\phi}(t,\vec{\theta})
    =
    \hat{q}
    +
    \frac{\hat{p}}{\kappa_D}\,f_D(t)
    +
    \hat\varphi(t,\vec{\theta})~.
\end{align}
Note, moreover, that the operators \(\hat{q},\hat{p}\) commute with all \(\hat a_{\ell\vec{m}},\hat a_{\ell\vec{m}}^{\dagger}\).

\paragraph{The full Hilbert space.}

The oscillator vacuum, i.e. the standard Fock vacuum for the $\ell \geq 1$ modes, is defined by
\begin{align}\label{eq:fock-vac}
    \hat a_{\ell\vec{m}}\ket{0}=0~,
    \qquad
    \ell\geq 1~.
\end{align}
One can build excited states over this vacuum by acting with creation operators $\hat{a}^\dagger_{\ell\vec{m}}$. The states built this way form the ordinary bosonic Fock space, $\cF_\t{osc}$. 

However, the full Hilbert space is spanned by states of the form 
\begin{align}
    \ket{\psi}_\t{QM} \otimes \hat{a}_{\ell_1 \vec{m}_1}^{\dagger} \hat{a}_{\ell_2 \vec{m}_2}^{\dagger}\cdots\ket{0}~,
\end{align}
where $\ket{\psi}_\t{QM}$ is an arbitrary state of the particle-on-a-ring Hilbert space, $L^2\qty\big(\S^1)$. We see that, akin to  the case of the non-compact scalar field \cite{Kirsten:1993ug}, the full Hilbert space $\cH $ is not a standard Fock space. In particular, we have
\begin{align}\label{eq:compact-scalar-hilbert-space-dsd}
    \cH = L^2\qty\big(S^1)\otimes\cF_\t{osc}~.
\end{align}
It is clear that this Hilbert space harbours infinitely many ``zero-particle'' states, namely
\begin{align} \label{family of zero-oscillator states}
    \ket{\Omega_n}\coloneqq \ket{n}_\t{QM} \otimes\ket{0}~,\qq{with} n=0, \pm 1, \pm 2,\ldots
\end{align}
and any linear combination thereof. Each $\ket{\Omega_n}$ is an eigenstate of $\hat{p}$ (see \cref{eigenstates of p}). We want to identify a subset of these states as vacua. If the only requirement is that the vacuum state is normalisable and has zero number of particles, then there are infinitely many vacua. However, there is another reasonable requirement that a candidate vacuum must satisfy: it must preserve dS isometries.%
\footnote{In a spacetime with a globally well-defined timelike Killing vector, there is another way to characterise the vacua: their energy. In static spacetimes the zero-particle states $\ket{\Omega_n}$  have energy proportional to $n^2$, with the proportionality constant being a non-negative number dictated by the geometry of the spatial slice, see e.g. \cite{Vitouladitis:2025zoy}. This procedure also selects only $\ket{\Omega_0}$ as the true vacuum state---except in the infinite-volume limit where spontaneous symmetry breaking predicts a manifold of vacua.}
As in the case of the non-compact scalar \cite{Kirsten:1993ug, Tolley:2001gg, Gibbons-Higuchi-Yang-Mills}, only the state that is annihilated by $\hat{p}$ is \ds-invariant, that is,
\begin{align} \label{unique dS-invariant zero-photon state}
    \ket{\Omega_0} = \ket{0}_\t{QM} \otimes \ket{0}~.
\end{align}
This is the unique \ds-invariant vacuum state in the full Hilbert space. Let us stress here that, contrary to the case of the non-compact scalar, the vacuum is normalisable. The states $\ket{\Omega_n}$ with $n \neq 0$ can be considered non-\ds-invariant zero-particle states (which are also normalisable). A detailed analysis of the dS (non-)invariance of these states is given in the following sections. 

Finally, one can also define a Fourier-transformed version of the zero-particle states:
\begin{align}\label{eq:vartheta-vac}
    \ket*{\tilde{\Omega}_\vartheta} \coloneqq \sum_{n\in\Z} \ex{-\ii n \vartheta} \ket{\Omega_n}~.
\end{align}
These states obviously still contain no particles. They diagonalise the exponentiated position operator, and are shuffled by the action of the exponentiated momentum operator:
\begin{align}\label{eq:Ovartheta-action}
    \ex{\ii \hat{q}}\ket*{\tilde{\Omega}_\vartheta} = \ex{\ii \vartheta} \ket*{\tilde{\Omega}_\vartheta}~, \qq{and} \ex{\ii\alpha\hat{p}}\ket*{\tilde{\Omega}_\vartheta} = \ket*{\tilde{\Omega}_{\vartheta-\alpha}}~.
\end{align}
In contrast to the momentum eigenstates, $\ket{\Omega_n}$, these states are not normalisable. Of course one can regulate them with a Gaussian profile, $\sim\ex{-\beta n^2}$, \cite{Benini:2025hbj} in the sum over $n$, which would, however, spoil their orthogonality. The flat-space counterparts of $\ket*{\tilde{\Omega}_\vartheta}$ are clustering symmetry-breaking vacua in $D>2$.


\subsection{\tps{\SO(D,1)}{SO(D,1)} UIRs  in the space of classical modes} 
\label{Subsec:UIRs-BD-modes}

One can infer the transformation properties of states under spacetime isometries in a variety of ways. A commonly employed method is to calculate a correlation function, over some state, of operators whose behaviour under isometries is known. The transformation of the correlation function then reveals information about the properties of the state. 
  
  Here, we consider a more direct approach.  First, we discuss the dS transformations of the classical Bunch--Davies modes. We then determine the quantum dS charges   and consider their action directly on the QFT Hilbert space. 
This discussion will be facilitated by first identifying the $\SO(D,1)$ unitary irreducible representations (UIRs) that are realised by our compact scalar field and its  classical $\ell \geq 1$ modes (and, thus, the oscillator single-particle Hilbert space). For a detailed representation-theoretic analysis on the full Hilbert space of the non-compact  massless scalar on global dS spacetime of any dimension see \cite{Gibbons-Higuchi-Yang-Mills}.

\paragraph{dS transformation of modes.}

Consider first an infinitesimal dS boost Killing vector, 
\begin{align}\label{dS boost specific}
    B_{\vec{0}} = B_{\vec{0}}^{\mu} \pd_{\mu} = \frac{1}{\rds}\qty(\cos{\theta_{D-1}}\, \frac{\partial}{\partial t} - \tanh{t} ~\sin{\theta_{D-1}}\, \frac{\partial}{\partial {\theta_{D-1}}} )~,
\end{align}
where $\vec{0} \equiv (0,\cdots,0)$, see \cref{def: boost Killing vectors}.
The classical modes transform as follows under this boost \cite{Higuchi:1986wu, Gibbons-Higuchi-Yang-Mills}. First, for $\ell\geq 2$ we have:
\begin{align}\label{dS-d transf of normalised modes ell>=2}
\lie_{B_{\vec{0}}}  \phi_{\ell \vec{m}} & \coloneqq \rds\, B_{\vec{0}}^{\mu} \pd_{\mu}   \phi_{\ell \vec{m}} = -\frac{i}{2} \cA_{\ell,\vec{m}}\,\phi_{\ell+1,\vec{m}} 
   -\frac{i}{2}\cA_{\ell-1,\vec{m}}\, \phi_{\ell-1,\vec{m}}~,
\end{align}
where, 
\begin{align}\label{eq:coeff-A-lm}
    \cA_{\ell,\vec{m}} &\coloneqq \sqrt{\frac{(\ell+ m_{D-2}+D-2)(\ell-m_{D-2}+1)\ell(\ell+D-1)}{(\ell+\frac{D-2}{2}) (\ell+\frac{D}{2})}  }~. 
\end{align}
Note that while we denote the coefficients as $\cA_{\ell,\vec{m}}$ they only depend on $\ell$ and on the first entry of the vector $\vec{m}$, namely $m_{D-2}$. For $\ell=1$ we have, instead 
\begin{align}\label{dS-d transf of nrmlsd modes ell=1}
\lie_{B_{\vec{0}}}  \phi_{1, \vec{m}} 
&=-\frac{i}{2}  \cA_{1,\vec{m}}\,\phi_{2,\vec{m}} 
   -\frac{i}{2}\tilde{\cA}_{0,\vec{m}}\, \frac{1}{\sqrt{c_0}}~,
\end{align}
where we defined 
\begin{align}
    \label{eq:tildeA1m}
    \tilde{\cA}_{0,\vec{m}} \coloneqq \sqrt{\frac{2(m_{D-2}+D-2)(1-m_{D-2})}{D(D-2)}}~,
\end{align}
and
\begin{align}
   \label{eq:c0}
   c_0 &\coloneqq \frac{\pi^{(D+1)/2}}{\Gamma\qty(\frac{D+1}{2})}~.
\end{align}
For $D>3$ we have $m_{D-2}\in \set{0,1}$, while for $D=3$ we have $m_{D-2}=m_1\in\set{0,\pm 1}$. The above formulas extend also to $D=2$ by letting $m_{D-2}=0$. Note, however, that in the $D=2$ case $\ell$ takes both positive and negative values. Further comments concerning the $D=2$ case will be given below.
The constant mode is trivially annihilated: 
\begin{align}\label{eq:dS-d transf of constant zero-mode}
\lie_{B_{\vec{0}}}  1= 0~,
\end{align}
while the time-dependent zero-mode $f_{D}(t)$ transforms into a linear combination of positive-frequency and negative-frequency modes as:
\begin{align}\label{eq:dS-d transf of time-dep zero-mode}
&\lie_{B_{\vec{0}}}  f_{D}(t) 
 =\frac{1}{\sqrt{2\,D\,c_0}} \left(   \phi^{\vphantom{*}}_{1,\vec{0}} +\phi_{1,\vec{0}}^{*} 
   \right).
\end{align}
In the above we have presented the expression for $\ell=1$ separately from the general case with $\ell \geq 2$. In particular, the explicit expression for $\lie_{B_{\vec{0}}}  \phi_{1, \vec{m}}$ cannot be found by simply letting $\ell=1$ in eq. \cref{dS-d transf of normalised modes ell>=2}. The reason is that the constant mode ``1'' on the right-hand side of \cref{dS-d transf of nrmlsd modes ell=1} has no normalisation factor, unlike the modes with $\ell \geq 1$ on the right-hand side of \cref{dS-d transf of normalised modes ell>=2}. Finally, the transformation properties of the negative frequency modes $\phi_{\ell \vec{m}}^{*}$ can be found by simply taking the complex conjugates of \cref{dS-d transf of normalised modes ell>=2,eq:coeff-A-lm,dS-d transf of nrmlsd modes ell=1,eq:tildeA1m,eq:c0}.

The transformation properties of the modes under $\so(D)$ rotations, i.e. under the Killing vectors of the spatial $S^{D-1}$, are determined completely from the well-known representation-theoretic properties of scalar spherical harmonics on $S^{D-1}$. See e.g. \cite{Higuchi:1986wu}. Finally, one can immediately obtain the transformation properties under the remaining $D-1$ dS boosts \cref{def: boost Killing vectors}, $B_{\vec{m}}$ (with $\vec{m} \neq 0$), from the infinitesimal $\so(D)$ transformations and the dS boost $\lie_{B_{\vec{0}}}$, as a consequence of the dS algebra.  

 \paragraph{UIR of  $\SO(D,1)$ in the space of Bunch--Davies modes.} 
 
 The positive (or negative) frequency Bunch--Davies modes, constructed above, form a UIR of $\SO(D,1)$ \cite{Higuchi:1986wu, Higuchi:1987hw}. The constant mode does not transform into any positive (or negative) frequency Bunch--Davies mode under any dS transformation. Instead, it is annihilated by all Lie derivatives. Thus, the constant mode, $1$, forms a \ds-invariant one-dimensional subspace. On the other hand, the Bunch--Davies modes $\phi_{1,\vec{0}}$ and $\phi^{*}_{1,\vec{0}}$ transform into the constant mode (see \cref{dS-d transf of nrmlsd modes ell=1}). However, since the constant mode has zero  norm and is also orthogonal to all Bunch--Davies modes with respect to the \ds-invariant Klein--Gordon inner product, we may identify the constant mode with zero in the Bunch--Davies solution space. To be specific, this is achieved by considering the ``physical'' solution space, i.e. the quotient vector space
 \begin{equation}\label{eq:VBD}
     V_\t{BD}  =  \spn{\phi_{\ell\vec{m}}}_{\ell\geq0}\Big/\spn{1}~.
 \end{equation}
Here, $\spn{\phi_{\ell m}}_{\ell\geq 0}$ is the space of all positive frequency Bunch--Davies modes with $\ell = 0,1,2,\ldots$, while, with slight abuse of notation, we let $\phi_{0\vec{0}}\equiv 1$. An analogous quotient space can be defined in the case of  negative frequency modes, $ V_\t{BD}^{*} \kern-1pt = \spn{\phi^{*}_{\ell\vec{m}}}_{\ell\geq0}\big/\kern-1pt\spn{1}$. The quotient space  $ V_\t{BD}$ (or  $ V_\t{BD}^{*}$), furnishes a unitary irreducible representation (UIR) of the dS group $\SO(D,1)$ \cite{Higuchi:1986wu} because the following criteria are satisfied:%
 \begin{enumerate}
     \item Positive definiteness of the norm.%
     \footnote{For positive-frequency modes, the positive-definite norm is given by the Klein--Gordon norm \cref{KG scalar product}. For negative-frequency modes, it is defined using the negative of the Klein--Gordon scalar product. Positive- and negative-frequency modes do not transform into each other under dS transformation and they separately form (equivalent) UIRs of $\SO(D,1)$.}
     \item Anti-hermiticity of dS generators \cref{dS-d invariance of KG scalar product}.
 \end{enumerate}
The representation that $V_\t{BD}$ furnishes depends on the spacetime dimension. We therefore separate the following cases.

\paragraph{$D \geq 4$.}

The corresponding representation belongs to the \emph{exceptional series} UIR of $\SO(D,1)$. In the notation of \cite{Penedones:2023uqc} this corresponds to the ``type I'' exceptional series with $\Delta = D-1$ and $s=0$. In the notation of \cite{Higuchi:1987hw, Higuchi:1986wu, Letsios:2023qzq, Hinterbichler:2026xqf}, this representation corresponds to the exceptional series UIR with  $F_{0} = -1$ and $F_1 = 1$. To make a comparison with \cite{Basile:2016aen},  $F_0$ corresponds to the negative of the scaling dimension and $F_1$ to the spin of the representation. The value $F_1=1$ refers to the fact that the representation is realised on the space of ``field strength'' solutions, $\set{\pd_\mu \phi_{\ell \vec{m}}}_{\ell \geq 1}$.

\paragraph{$D = 3$.} The corresponding representation belongs to the principal series UIR of $\SO(3,1)$ with $\Delta=1$, $s=1$ \cite{Penedones:2023uqc} (or equivalently $F_0=-1$, $F_1=1$ \cite{Higuchi:1987hw, Higuchi:1986wu, Letsios:2023qzq, Hinterbichler:2026xqf}). This is the photon UIR of $\SO(3,1)$. Indeed, an alternative construction of the same UIR is achieved using the Bunch--Davies mode solutions of the massless spin-1 Maxwell gauge potential on $\ds_3$ in the Lorenz gauge:
\begin{align}\label{Maxwell equations dS_3}
    \Box A_{\mu} = \frac{2}{\rds^2}A_{\mu}~, \qquad \nabla^{\mu} A_{\mu}=0~.
\end{align}
Here, again, one has to consider the quotient of the solution space with the space of pure-gauge solutions \cite{Higuchi:1986wu, Higuchi:1987hw}. This fact, that on $\ds_3$ the physical photon modes and the physical compact scalar modes form the same $\SO(3,1)$ UIR, is a manifestation of the duality of interest. This is discussed further in \cref{sec:EM-Hilbert}.


\paragraph{$D = 2$.} In this case, the representation corresponds to the direct sum of two discrete series UIRs of $\SO(2,1)$ with $\Delta=1$ \cite{Anninos:2023lin,  Epstein:2014jaa, Farnsworth:2024yeh}. Now, the only angular momentum quantum number is the $\SO(2)$ quantum number $\ell$ which takes values in $\mathbb{Z}$. The transformation properties of the modes are found by letting  $m_{D-2}=0$ and $D=2$ in eqs. \cref{dS-d transf of normalised modes ell>=2}--\cref{eq:dS-d transf of time-dep zero-mode}. We have a direct sum  of UIRs in the space of Bunch--Davies modes because the two sets, $\set{ \phi_{\ell}}_{\ell \ \geq 1 }$ and $\set{ \phi_{\ell}}_{\ell \leq -1  }$, do not mix with each other under $\SO(2,1)$, while $\phi_{0} = 1$ is identified with zero.


\subsection{Classical dS charges}\label{subsec:classical-ds-charges}

The energy-momentum tensor of $\phi$ is found in the standard way by varying the action with respect to the metric. It reads:  
\begin{align}\label{def: energy-momentum tensor}
   {T^{\mu\nu}} =\kappa_D \rds^{2}  \qty( \partial^{\mu}\phi\, \partial^{\nu}\phi - \frac{1}{2}g^{\mu\nu}  \partial^{\alpha}\phi\, \pd_{\alpha}\phi )~,
\end{align}
with $T_{\mu \nu} = T_{\nu \mu}$ and $\nabla^{\mu}T_{\mu \nu}=0$. For convenience, the tensor $T^{\mu \nu}$ in \cref{def: energy-momentum tensor} is defined to be dimensionless. The physical energy-momentum tensor is obtained by multiplying $T^{ \mu \nu}$ with $(\rds)^{-D}$.

 An interesting feature of the minimally coupled massless scalar field (compact or non-compact) is that its dS charges receive contributions both from the oscillator and from the zero-mode sectors, unlike a generic massive scalar field with positive mass-squared.  In particular, decomposing the field into its  zero-mode and oscillator parts, $\phi=\phi_0 + \varphi$, we find  that the energy-momentum tensor can be split into three parts that are separately symmetric and conserved:
\begin{align}\label{split of stress tensr in 3 parts}
    T^{\mu\nu} = T^{\mu\nu}_{(\text{osc})}+T^{\mu \nu}_{(\text{top})} + T^{\mu\nu}_{(\text{cross})},
\end{align}
 where\footnote{Note that, in our conventions, $\pd_t = - \partial^t = \rds^{-1}~\partial / \partial t$.}
 \begin{align}
   \frac{1}{\kappa_D \rds^{2}}\, T_{(\text{osc})}^{\mu\nu} 
   &= 
   \partial^{\mu}\varphi\, \partial^{\nu}\varphi 
   - \frac{1}{2}g^{\mu\nu}  \partial^{\alpha}\varphi\, \pd_{\alpha}\varphi~, \nonumber \\
   \frac{1}{\kappa_D \rds^{2}}\,  T^{\mu\nu}_{(\text{top})} 
   &=
    \partial^{\mu}\phi_{0}\, \partial^{\nu}\phi_{0} 
    - \frac{1}{2}g^{\mu\nu}  \partial^{t}\phi_{0}\, \pd_{t}\phi_{0}~, \nonumber\\
   \frac{1}{\kappa_D \rds^{2}}\, T^{\mu\nu}_{\text{(cross)}} 
   &= 
   \partial^{\mu}\varphi\, \partial^{\nu}\phi_{0}+ \partial^{\mu}\phi_{0}\, \partial^{\nu}\varphi 
   - g^{\mu\nu}  \partial^{\alpha}\varphi\, \pd_{\alpha}\phi_{0} . 
\end{align}

\noindent  \textbf{dS charges.} The standard definition for the conserved  currents associated with dS isometries is 
\begin{align}
    J^{\mu}_{(\xi)} = T^{\mu}_{\phantom{\mu}\nu}\, \xi^{\nu}~,
\end{align}
where $\xi^{\mu}$ is a Killing vector satisfying $\nabla_{(\mu} \xi_{\nu)}=0$. It is easy to check that $\nabla^{\mu} J_{(\xi)\mu}=0$. There is one dS current $ J^{\mu}_{(\xi)}$ for each Killing vector. The time-independent dS charges associated with dS Killing vectors $\xi^{\mu}$ are
\begin{align}
    Q^{\ds}{(\xi)} = \int_{S^{D-1}}  \dd[D-1]{\theta} \sqrt{-g}\; \no{\kern1pt J^{t}_{(\xi)}} = \int_{S^{D-1}}  \dd[D-1]{\theta} \sqrt{-g}\;  \no{\kern1pt T^{t}_{\phantom{t}\nu}\,\xi^{\nu}}~.
\end{align}
We use the symbol $\no{\,\cdots\,}$ to denote normal ordering (placing all creation operators to the left of annihilation operators), which will be relevant for the quantum theory.

  It is clear that  each dS charge, associated with a Killing vector $\xi^{\mu}$, can also be split into three parts as:
  \begin{align}\label{dS-d charges split into 3 parts}
      Q^{\ds}(\xi) =  Q^{\ds}_{(\text{osc})}(\xi) + Q^{\ds}_{(\text{top})}(\xi)+Q^{\ds}_{(\text{cross})}(\xi),
  \end{align}
where
 \begin{align}
       Q^{\ds}_{(\text{A})}(\xi) &= \int_{S^{D-1}}  \dd[D-1]{\theta} \sqrt{-g}\; \no{\kern1pt T_{(\text{A})}^{t \nu}\,\xi_{\nu}}~, \qquad \text{A} \in \set{ \text{osc}, \text{top}, \text{cross}}.
      \end{align} 
Each of these charges is separately conserved in time.

\subsection{Quantum dS charges}

We now come to the quantum version of the dS charges, and discuss their action on the Hilbert space.
\paragraph{The oscillator  contribution.} This contribution to the dS charge, \cref{dS-d charges split into 3 parts}, is the standard one and can be conveniently expressed as (see, e.g. \cite{Higuchi:1987hw}):
\begin{align}\label{dS charge osillator in terms of KG general}
 \hat{Q}^{\ds}_{(\text{osc})}(\xi)
       &= -\frac{\ii\,\kappa_D}{2}\int_{S^{D-1}} \dd[D-1]{\theta} \sqrt{-g}\; \no{J^{t}_\t{KG}\qty(\hat{\varphi}, \lie_{\xi} \hat{\varphi})}  \\
       &= -\frac{\ii\,\kappa_D}{2} \no{\kip{\hat{\varphi}}{\lie_{\xi}\hat{\varphi}}}~.
\end{align}
This expression relates the oscillator dS charge with the Klein--Gordon current \cref{def: KG current dS-d} and inner product \cref{KG scalar product}. Note that the dS invariance of the Klein--Gordon inner product \cref{dS-d invariance of KG scalar product} makes manifest the hermiticity of  $\hat{Q}^{\ds}_{(\text{osc})}(\xi)$.
The charge $\hat{Q}^{\ds}_{(\text{osc})}(\xi)$ generates the single-particle UIR of $\SO(D,1)$ in the oscillator sector of the theory  by acting on single-particle states $\hat{a}^{\dagger}_{\ell \vec{m}} \ket{0}$, as in the case of the non-compact scalar field \cite{Gibbons-Higuchi-Yang-Mills}. For completeness, we review  in \cref{App: single particle UIRs in Hilbert space} how this can be readily shown from the representation-theoretic properties of the Bunch--Davies modes given in eqs. \cref{dS-d transf of normalised modes ell>=2}--\cref{eq:dS-d transf of constant zero-mode}.
Note that the Fock vacuum $\ket{0}$ \cref{eq:fock-vac} is annihilated by all oscillator dS charges $\hat{Q}^{\ds}_{(\text{osc})}(\xi)$, i.e. for any dS Killing vector $\xi^{\mu}$  (see \cref{App: single particle UIRs in Hilbert space}).


\paragraph{The topological contribution.} The purely topological part of the dS charge \cref{dS-d charges split into 3 parts} is
\begin{align}
   \hat{Q}^{\ds}_{(\text{top})}(\xi)
    & =   - \frac{\kappa_D \rds^2}{2} \int_{S^{D-1}} \dd^{D-1}\theta \sqrt{-g}\;  \xi^{t}\qty(\pd_t\hat{\phi}_0(t))^2   \nonumber \\
   & = - \frac{\hat{p}^2}{2\kappa_D\,\Vol_{D-1}^2}\;\frac{1}{\cosh^{D-1}{t}} \int_{S^{D-1}} \dvol_{D-1}\,\xi^{t}~,
\end{align}
where we used that
\begin{equation}
\pd_t\hat{\phi}_0(t)
    =
    \frac{\hat{p}}{\kappa_D\,\rds \Vol_{D-1}}\, \frac{1}{\cosh^{D-1}{t}}~,
\end{equation}
as well as $\pd_i\hat{\phi}_0(t) = 0$. It follows that $\hat{Q}^{\ds}_{(\text{top})}(\xi)$ vanishes for all Killing vectors. Let us discuss rotational and boost Killing vectors separately. For a rotational Killing vector, $\xi^\mu = R^{\mu}$, the proof is trivial as $R^{t}=0$, and thus $\hat{Q}^{\ds}_{(\text{top})}(R)=0$. For a boost Killing vector, $\xi^\mu = B^{\mu}_{\vec{m}}$, we have from \cref{def: boost Killing vectors} that $B_{\vec{m}}^t = Y_{\ell=1,\vec{m}}(\vec{\theta})$ and so 
\begin{align}
   \hat{Q}^{\ds}_{(\text{top})}\qty\big(B_{\vec{m}}) &\propto \int_{S^{D-1}} \dd[D-1]{\theta} \sqrt{\tilde{g}}\; Y_{\ell=1,\vec{m}}(\vec{\theta}) = 0~,
\end{align}
by orthogonality with the constant spherical harmonic $Y_{0\vec{0}} = \qty(\Vol_{D-1})^{-1/2}$. In conclusion,
\begin{equation}
    \hat{Q}^\ds_\t{(top)}(\xi) = 0~,
\end{equation}
for all Killing vectors $\xi^\mu$.

\paragraph{The cross contribution.} A short calculation reveals that the cross contribution to the dS charge \cref{dS-d charges split into 3 parts} is 
\begin{align}\label{cross contribution in dS charge}
  \hat{Q}^{\ds}_{(\text{cross})}(\xi)
  &= -\frac{\hat{p}}{\Vol_{D-1}}\int_{S^{D-1}}  \dvol_{D-1} \lie_{\xi}\hat{\varphi}~,
\end{align}
where we recall that $\lie_{\xi} \hat{\varphi} = \rds\, \xi^{\nu}\pd_{\nu}\hat{\varphi}$. This integral is non-vanishing only if $\lie_{\xi}\hat{\varphi}$ contains a constant ($\ell =0$) term in its mode expansion. It can be readily shown that this does not happen if $\xi^{\mu}$ is a rotational Killing vector $R^{\mu}$ since $\SO(D)$ rotations keep $\ell$ fixed. Thus 
\begin{align}
    \hat{Q}^{\ds}_{(\text{cross})}(R)=0.
\end{align}
However, for a boost Killing vector, $B^\mu$, the integrand contains a constant term, due to the transformation of the mode $\phi_{1, \vec{m}}$ in \cref{dS-d transf of nrmlsd modes ell=1}. Thus, $ \hat{Q}^{\ds}_{(\text{cross})}(B) \neq 0$. It is sufficient to show this for one of the $D$ boosts; we choose $B=B_{\vec{0}}$ \cref{dS boost specific}.  We find
\begin{align}\label{cross contribution in dS charge B_0}
 & \hat{Q}^{\ds}_{(\text{cross})}\qty(B_{\vec{0}})
    = \frac{\ii}{\sqrt{2 D\, c_0\, \kappa_D}}\; \hat{p}\,\qty(\hat{a}_{1,\vec{0}} - \hat{a}_{1,\vec{0}}^{\dagger})~,
\end{align}
where we used \cref{dS-d transf of nrmlsd modes ell=1}. Note also that $\hat{Q}^{\ds}_{(\text{cross})}(B_{\vec{0}})$ is hermitian.

\paragraph{Summary of dS charges.} To sum up, the full  quantum dS charges \cref{dS-d charges split into 3 parts} are hermitian and have the following explicit contributions:
\begin{itemize}
    \item \textbf{Rotations:} The only contribution is the oscillator one
 \begin{align}
 \hat{Q}^{\ds}( R)=\hat{Q}^{\ds}_{(\text{osc})}(R)~, \qquad \text{for}\ R^{\mu} \in \so(D)~,
\end{align}
where $\hat{Q}^{\ds}_{(\text{osc})}(R)$ is given in eqs. \cref{dS charge osillator in terms of KG general,dS-d charge osillator in terms of KG explicit}. Note that 
\begin{align}\label{eq:Q(R) on fock vac}
    \hat{Q}^{\ds}_{(\text{osc})}(R) \ket{0}=0~,    
\end{align}
and thus, $ \hat{Q}^{\ds}(R) \ket{\Omega_n}=0$.

\item \textbf{Boosts:} There are two contributions, one from the oscillators and the cross contribution:
 \begin{align}\label{eq:Q nonzero rots and boosts}
 \hat{Q}^{\ds}(B)&=\hat{Q}^{\ds}_{(\text{osc})}(B)+\hat{Q}^{\ds}_{(\text{cross})}(B)~,\qquad\text{for}\ B^{\mu}\ \text{a dS boost.}
\end{align}
The explicit expression for $\hat{Q}^{\ds}_{(\text{osc})}(B)$ is given in eqs. \cref{dS charge osillator in terms of KG general}, \cref{dS-d charge osillator in terms of KG explicit}. Note that again 
\begin{align}\label{eq:Q(B) on fock vac}
    \hat{Q}^{\ds}_{(\text{osc})}(B) \ket{0}=0~.    
\end{align}
The cross contribution, $\hat{Q}^{\ds}_{(\text{cross})}(B)$, is defined in \cref{cross contribution in dS charge}, and more explicitly for $B_{\vec{0}}^{\mu}$ in \cref{cross contribution in dS charge B_0}. It remains to check whether the candidate vacua, $\ket{\Omega_n}$, are invariant under $\hat{Q}^{\ds}_{(\text{cross})}(B)$.

\end{itemize}


\paragraph{Quid est vacuum?} Having analysed all contributions to dS charges, we would like to understand which of the zero-particle states $\ket{\Omega_n}=\ket{n}_\t{QM}\otimes\ket{0}$ are \ds-invariant. First, note that all of these states are $\SO(D)$-invariant. Indeed, by \cref{eq:Q(R) on fock vac} they are all annihilated by rotational charges $\hat{Q}^{\ds}(R)$. This part is identical to the non-compact scalar case \cite{Kirsten:1993ug}.

We now turn to boosts. It is sufficient to focus our attention to one of the boost generators, $B_{\vec{0}}$. Armed with the explicit expression of $\hat{Q}^{\ds}(B_{\vec{0}})$ we can directly act on $\ket{\Omega_n}$. Of the two non-vanishing contributions of $\hat{Q}^{\ds}(B_{\vec{0}})$, from \cref{eq:Q nonzero rots and boosts} we have already shown that the oscillator part annihilates all zero-particle states. Therefore: 
\begin{align}
       \hat{Q}^{\ds}(B_{\vec{0}})\ket{\Omega_n} &=   \hat{Q}^{\ds}_{(\text{cross})}(B_{\vec{0}})\ket{\Omega_n}~.
\end{align}
Using the explicit form of the charge \cref{cross contribution in dS charge B_0}, as well as \cref{eigenstates of p}, we find:
\begin{align}
       \hat{Q}^{\ds}_{(\text{cross})}(B_{\vec{0}}) \ket{\Omega_n} = -\frac{\ii\, n}{\sqrt{2 D\, c_0\, \kappa_D}}\, \hat{a}^\dagger_{1,\vec{0}} \ket{\Omega_n}~,
\end{align}
from which we see that only the uncharged state, $n=0$, is invariant\footnote{The situation is similar for the other boosts, $B_{\vec{m}}$, as follows from the dS algebra and \cref{App: single particle UIRs in Hilbert space}.}. In particular, the action of $\hat{Q}^{\ds}_{(\text{cross})}(B_{\vec{0}})$ on any zero-particle state, $\ket{\Omega_n}$, results in a one-particle state, with coefficient proportional to $n$. Thus, among these zero-particle states, only $\ket{\Omega_0} = \ket{0}_\t{QM}\otimes\ket{0}$ is \ds-invariant. This is the unique state that is invariant under both dS isometries, and the internal shift symmetry of the compact scalar: 
\begin{align}
    \hat{Q}^\ds(\xi) \ket{\Omega_0} = 0 \qquad\t{and}\qquad \hat{p}\ket{\Omega_0} = 0~,~~~\text{for any Killing vector}~\xi^\mu.
\end{align}
 The situation is similar to the case of the non-compact scalar \cite{Kirsten:1993ug, Tolley:2001gg}, although there, the unique \ds-invariant zero-particle state is not normalisable.

\paragraph{Decompactification limit.} Let us make a comment on the non-compact case, en passant (see also \cref{app:limits}). To this end it will be convenient to rescale our zero-mode operators as $\hat{X} = \sqrt{\kappa_D}\, \hat{q}$ and $\hat{P} = \hat{p}/\!\sqrt{\kappa_D}$. This canonical pair still generates the quantum mechanics of the particle on a ring, only this time, the ring has radius $\sqrt{\kappa_D} \sim \rds^{(D-2)/2}\, f$. From here one can immediately read off the limit in which the scalar decompactifies
\begin{align}
    \sqrt{\kappa_D} \to \infty~.
\end{align}
In this limit our field becomes the usual minimally coupled massless scalar field. Accordingly, the quantum mechanical Hilbert space is that of a particle on an infinitely large ring, i.e. the real line: $L^2(\R)$. Tensored with the Fock space, one recovers the Hilbert space found by Kirsten and Garriga \cite{Kirsten:1993ug}, namely $\cH = L^2(\R)\otimes \cF_{\t{osc}}$. The expressions for the dS charges can be immediately read off from the ones we derived above. For instance, $\hat{Q}^{\ds}_{(\text{cross})}(B_{\vec{0}})$ \cref{cross contribution in dS charge B_0} reads:
\begin{align}
    \hat{Q}^{\ds}_{(\text{cross})}(B_{\vec{0}}) = \frac{\ii}{\sqrt{2 D\, c_0}}\; \hat{P}\,\qty(\hat{a}_{1,\vec{0}} - \hat{a}_{1,\vec{0}}^{\dagger})~,
\end{align}
where now the momentum operator has a continuous spectrum of eigenvalues. Once again, the true vacuum is the unique state satisfying
\begin{align}
    \hat{Q}^\ds(\xi) \ket{\Omega_0} = 0 \qquad\t{and}\qquad \hat{P}\ket{\Omega_0} = 0~,
\end{align}
for all Killing vectors $\xi^\mu$, thus providing a different derivation of the conclusion of \cite{Kirsten:1993ug}.


\subsection{Observer-dependent particle number?}

In this subsection, we present a simple but interesting phenomenon that occurs in the case of the massless minimally coupled scalar (at least within our quantisation scheme), and has no analogue in the  free QFT of a massive ($m^2>0$) scalar  on $\ds_D$ equipped with a \ds-invariant vacuum. 

As we have explained above, the massless scalar has many normalisable states, $\ket{\Omega_n}$, with zero oscillator quanta. One may call them ``zero-particle states''. Interestingly, the notion of a zero-particle state is not common to all geodesic observers as described from their respective frames. Whenever observers have access to sectors of the Hilbert space with a non-zero shift charge $n$,  some of the de~Sitter transformations appear to ``create particles''.  In  these sectors, the particle number operator is not invariant under dS boosts, and thus boosts mix states with different oscillator particle number. This does not happen for a \ds\ invariant zero-particle state, such as $\ket{\Omega_0}$. In that case, all geodesic observers agree that $\ket{\Omega_0}$ has zero particles, much as inertial observers in Minkowski spacetime agree on  the Poincaré-invariant vacuum state. Note that infinitely many zero-particle states also exist in the case of the non-compact massless scalar, but they are not normalisable \cite{Kirsten:1993ug}, including the unique \ds-invariant one. By contrast, in our case, the zero-particle states $\ket{\Omega_n}$ are normalisable,  and the field-strength two-point functions computed in these states are Hadamard (see \cref{subsec: restoration of dS symmetry 2-point}). This enables us to draw physically meaningful conclusions.

Timelike geodesics in de~Sitter spacetime transform among themselves under dS transformations.%
\footnote{See  \cite{Taylor:2024vdc} for recent discussions on scattering in global dS spacetime with special emphasis on geodesic observers.}
Consider a geodesic observer on global $\ds_D$ who performs measurements with some kind of device that can detect de~Sitterian quanta. In  $\ds_3$ where the compact scalar is dual to electromagnetism, one can think of this as an ideal cosmological photodetector, which ticks when a photon is absorbed.  Suppose the field is prepared in the state $\ket{\Omega_0} = \ket{0}_\t{QM} \otimes \ket{0}$. The detector registers no excitations, with probability 1.
 Since $\ket{\Omega_0}$ is \ds-invariant, all geodesic observers with identical local detectors ---  related to one another by dS transformations --- agree that the state $\ket{\Omega_0}$ has zero particles. More generally, all such observers agree on the particle content of the states, since any dS transformation that transforms one geodesic to another acts on the Hilbert space while preserving particle number.  Single-particle states transform into linear combinations of single-particle states, two-particle states into linear combinations of two-particle states and so forth. In other words, each  $k$-particle oscillator Hilbert space with $n=0$ furnishes a unitary representation of $\SO(D,1)$.%
\footnote{For the decomposition of multi-particle Hilbert space of dS QFTs into a direct sum of UIRs of $\SO(D,1)$ for various theories see \cite{Penedones:2023uqc}.}

Consider now an observer whose state belongs to a sector of the Hilbert space with $n\neq 0$. A transition from the $n=0$ sector to an $n \neq 0$ sector can be achieved by acting with the operator $\ex{i n \hat{q}}$ ($n=\pm 1, \pm 2,\cdots$) on $\ket{\Omega_0}$:
\begin{align}
\ex{i n \hat{q}} \ket{\Omega_0} = \ket{\Omega_n}~.
\end{align}
In $\ds_3$, from the electromagnetic point of view, this creates a magnetic monopole with magnetic charge $n$. Excited states in the $n \neq 0$ sector are created by acting with $\hat{a}^{\dagger}_{\ell \vec{m}}$ ($\ell \geq 1$) on the zero-particle state $\ket{\Omega_n}$. Interestingly, although our first geodesic observer would detect no oscillator quanta when the field is prepared in the state $\ket{\Omega_n}$, this notion of  zero-particle state is not common to all geodesic observers with access to the same $n \neq 0$ sector. 

To see this explicitly, take the second geodesic observer to be related to the first one by an infinitesimal dS boost, say $B_{\vec{0}}$ in \cref{dS boost specific}. In the boosted frame, the state of the system is 
\begin{align}
   \qty(\hat{1}+\ii\,\epsilon \,\hat{Q}^{\ds}(B_{\vec{0}})) \ket{\Omega_n}, \qquad \abs{\epsilon} \ll 1~.
\end{align}
This state contains a non-vanishing one-particle component with $(\ell ,\vec{m})=(1, \vec{0})$ because of the  cross term in \cref{cross contribution in dS charge B_0}. The word ``particle'' should be treated with extra care:%
\footnote{More than the usual, cf. \cref{eq:particles?}.}
the $\ell=1$ quanta  have very large wavelengths and may not be detectable in practice%
\footnote{Such ``particles'' correspond to the eigenvalue $-\ell(\ell+D-2)\big|_{\ell=1}=-(D-1)$ of the spatial Laplacian. At $t=0$, their wavelength is $\lambda \sim \frac{2\pi}{\sqrt{D-1}} \rds$. In four dimensions this wavelength is approximately 3.63 Hubble radii.}
Nevertheless, the two observers do not agree on the particle content of the state because, in a charged sector, \ds\ boosts  mix states with different oscillator particle numbers. 

For a finite boost, with parameter $T \in \mathbb{R}$, the relevant state in the boosted frame is 
\begin{align}
    \exp(\ii\,T\,\hat{Q}^{\ds}(B_{\vec{0}}))\ket{\Omega_n}~.
\end{align}
Such a transformation gives rise to a linear combination of states with different angular momenta $\ell$. Some of these states will have sufficiently large $\ell$ to be visible to a local detector. It would be interesting to connect this phenomenon with recent work on measurements in QFT \cite{Anastopoulos2022QFTQI, Anastopoulos2023TowardsFTRQI, Fewster:2023cfq, Fewster:2024pur, Bostelmann:2020unl}.

We expect that a variation of this effect should persist in certain interacting theories. Suppose, for example, that an axion potential is generated as in \cref{eq:action-compact}, and that this potential preserves a subgroup of the shift symmetry, for instance $v(\phi)\sim \cos(2\phi)$, preserving $\Z_2\subset \U(1)$. The Hilbert space would then still decompose into sectors of fixed charge of the preserved symmetry. Then, while generally the notion of a fixed number of particles is disrupted, at weak coupling one can still talk about quasiparticles. The neutral vacuum state, defined with respect to the corresponding quasiparticle creation and annihilation operators, will presumably remain \ds\ invariant. An argument for this follows by the definition of the Euclidean vacuum via path integral methods, see \cref{sec:sphere}. In contrast, charged zero-quasiparticle states would still fail to be \ds\ invariant; indeed it would be remarkable if they were to become invariant. Then, the quasiparticle content of charged sectors would change under \ds\ transformations, providing an interacting version of the effect described above. If such quasiparticles are sufficiently long-lived, it might be detectable by our inertial observers.

\paragraph{$\SO(D,1)$ representation on the full Hilbert space.} From the mathematical viewpoint, we note that in sectors with $n \neq 0$, a $k$-particle state
\begin{equation}
     \hat{a}_{\ell_1 \vec{m}_1}^{\dagger}\!\cdots\hat{a}_{\ell_k \vec{m}_k}^{\dagger}\ket{\Omega_n}~, \qquad n \neq0~,
\end{equation}
transforms into a linear combination of $(k+1)$- and $(k-1)$-particle states under infinitesimal dS boosts, see \cref{cross contribution in dS charge B_0}. Hence, boosts do not preserve particle number, while rotations do. Thus, the $k$-particle subspaces are not invariant under the action of $\SO(D,1)$. Nevertheless, for each fixed value of $n\neq0$, the full multi-particle Hilbert space, including all $k = 0,1,2,\ldots$, carries a unitary representation of $\SO(D,1)$: generators are hermitian and the norm is positive. It would be interesting to determine the decomposition of this representation into unitary irreducible representations of $\SO(D,1)$.


\subsection{Two-point functions in charged sectors and symmetry restoration}\label{subsec: restoration of dS symmetry 2-point}

The field strength of our compact scalar field is
\begin{align}
    \hat{J}_{\mu}(t,\bm{\theta}) = \partial_{\mu}\hat{\varphi}(t,\bm{\theta}) + \partial_{\mu}\hat{\phi}_{0}(t).
\end{align}
We now study field-strength Wightman two-point functions in the state $\ket{\Omega_n} = \ket{n}_{\t{QM}} \otimes \ket{0}$,
\begin{align}
    G^{(n)}_{\mu\nu'}\qty(x,x') \coloneqq \bra{\Omega_n}  \hat{J}_{\mu}(t,\bm{\theta}) \hat{J}_{\nu'}(t',\bm{\theta}') \ket{\Omega_{n}}.
\end{align}
Unprimed indices refer to the tangent space at $x=(t,\bm{\theta})$ while primed ones refer to the tangent space at $x'=(t',\bm{\theta}')$. The two-point function is split in four parts, namely
\begin{equation}
\begin{split}
  G^{(n)}_{\mu\nu'}(x,x') 
  &= 
  \mel{0}{\partial_{\mu}\hat{\varphi}(t,\bm{\theta})}{0} \; _{\t{QM}}\!\mel{n}{ \partial_{\nu'}\hat{\phi}_{0}(t')}{n}_{\t{QM}} 
  \nn 
  &\phantom{=~}+\; _{\t{QM}}\!\mel{n}{\partial_{\mu}\hat{\phi}_{0}(t)}{n}_{\t{QM}} \; \mel{0}{ \partial_{\nu'}\hat{\varphi}(t',\bm{\theta}')}{0}  \nn 
  &\phantom{=~}+\mel{0}{\partial_{\mu}\hat{\varphi}(t,\bm{\theta}) ~\partial_{\nu'}\hat{\varphi}(t',\bm{\theta}')}{0} \nn 
  &\phantom{=~}+\; _{\t{QM}}\!\mel{n}{\partial_{\mu}\hat{\phi}_{0}(t)~\partial_{\nu'}\hat{\phi}_{0}(t')}{n}_{\t{QM}}.
\end{split}
\end{equation}
Observing that $\mel{0}{\partial_{\mu}\hat{\varphi}(t,\bm{\theta})}{0}=\mel{0}{ \partial_{\nu'}\hat{\varphi}(t',\bm{\theta}')}{0}=0$, and using the explicit expression for the zero-mode operator \cref{quantised zero mode,time-dep zero-mode dS-d}, the two-point function reads
\begin{align}\label{field-strengt 2-point function any n}
  G^{(n)}_{\mu\nu'}(x,x') 
  &=
  \mel{0}{\partial_{\mu}\hat{\varphi}(t,\bm{\theta})\,\partial_{\nu'}\hat{\varphi}\qty(t',\bm{\theta}')}{0} + \frac{n^2}{\kappa_D^2\,\rds^2\,(\Vol_{D-1})^2}\,\frac{g_{\mu t}}{\cosh^{D-1}{t}}\,\frac{g_{\nu' t'}}{\cosh^{D-1}{t'}}.
\end{align}
Hence, the two-point function has two distinct parts: an oscillator and a zero-mode part. In the  oscillator part, the short-wavelength behaviour of the Bunch--Davies modes \cref{positive freq condition ell>>1} indicates that we should adopt a $-\ii\epsilon$ prescription, i.e. $t$ should be understood as $t-\ii \epsilon$ (with $0< \epsilon \ll 1$).

\paragraph{The $n=0$ case.} Among the field-strength two-point functions above, the $n=0$ case is special. It is the unique value of $n$ that renders the two-point function \ds-invariant, as a consequence of the dS invariance of 
$\ket{\Omega_{0}}$. In particular, the $n=0$ two-point function
\begin{align}
  G^{(0)}_{\mu\nu'}\qty(x,x') = \mel{\Omega_0}{\hat{J}_{\mu}\qty(t,\bm{\theta})\, \hat{J}_{\nu'}\qty(t',\bm{\theta}')}{\Omega_{0}}  
  =
  \mel{0}{\partial_{\mu}\hat{\varphi}(t,\bm{\theta})\,\partial_{\nu'}\hat{\varphi}(t',\bm{\theta}')}{0},
\end{align}
is \ds-invariant, as can be seen from its  infinitesimal dS transformation
\begin{align}
    \delta_\xi G^{(0)}_{\mu\nu'}\qty(x,x')
    &=
    \mel{0}{\comm{\partial_{\mu}\hat{\varphi}(t,\bm{\theta})}{\hat{Q}^{\ds}(\xi)}\, \partial_{\nu'}\hat{\varphi}\qty(t',\bm{\theta}')}{0} \nonumber \\
  &\phantom{=~} + \mel{0}{\partial_{\mu}\hat{\varphi}(t,\bm{\theta})\,\comm{\hat{Q}^{\ds}(\xi)}{\partial_{\nu'}\hat{\varphi}\qty(t',\bm{\theta}')}}{0}=0~.
\end{align}
The two-point function $G^{(0)}_{\mu\nu'}\qty(x,x')$ is infrared finite and has the Hadamard form in the limit $x=(t,\bm{\theta}) \to x'=\qty(t',\bm{\theta}')$ \cite{Allen:1987tz}. 

An explicit form for this two-point function can be found starting from the two-point function of a massive scalar $\ev*{\hat{\Psi}(t,\bm{\theta}) \hat{\Psi}\qty(t',\bm{\theta}')}$. This is both Hadamard and \ds-invariant \cite{Allen:1985wd}. $G^{(0)}_{\mu\nu'}(x,x')$ is retrieved from the massless limit of $\partial_{\mu}\partial_{\nu'}\ev*{\hat{\Psi}(t,\bm{\theta}) \hat{\Psi}\qty(t',\bm{\theta}')}$. 


\paragraph{The $n\neq 0$ sectors.} 

First, let us note that the singularity structure of the two-point function \cref{field-strengt 2-point function any n} is  determined only from the oscillator two-point function. Since the oscillator two-point function is Hadamard (as explained earlier), we conclude that the full two-point function $G^{(n)}_{\mu\nu'}(x,x')$ is also Hadamard for any $n \in \mathbb{Z}$.

However, for $n \neq 0$, the two-point function $G^{(n)}_{\mu\nu'}(x,x')$ is invariant under $\SO(D)$ but not under the full de~Sitter group. Interestingly, when $t$ or $t'$ tends to $\pm \infty$, the zero-mode contribution in \cref{field-strengt 2-point function any n} becomes negligible, as it decays exponentially, while the oscillator contribution decays only polynomially. Thus, in this limit, we have
\begin{align}
    G^{(n)}_{\mu\nu'}(x,x') \approx G^{(0)}_{\mu\nu'}(x,x')~, \qquad
 \text{for}~t \to \pm \infty~, ~\text{or}~t' \to \pm \infty~.
 \end{align}
Hence, the two-point function in a charged ``zero-particle'' state restores full $\SO(D,1)$ invariance at very late (and very early) times. 

\subsection{Comments on spontaneous symmetry breaking} 

As detailed in the introduction, the theory we consider arises in many cases as a theory of Goldstones, which would indicate spontaneous symmetry breaking (SSB). This is a delicate subject in de~Sitter for two reasons. First, due to the compactness of its spatial slices at finite global time,  and second, due to enhancement of infrared effects similarly to flat space QFT below the critical dimension \cite{Ratra:1985,Ford:1986,Boyanovsky:2012qs,DiPietro:2023inn}. A caveat to this is SSB of $(D-1)$-form symmetries, where the theory develops special states known as ``(p)universes'' \cite{Hellerman:2006zs,Sharpe:2019ddn,Aminov:2019hwg,Tanizaki:2019rbk,Komargodski:2020mxz}, allowing SSB in compact volume or low dimensions. This possibility was recently explored in $\ds_2$ \cite{Aguilera-Damia:2026dbk}.

In our case, the theory we consider does not exhibit spontaneous symmetry breaking (SSB) of its $\U(1)$ shift symmetry \cite{Ratra:1985}, unless it couples to discrete gauge fields \cite{Aguilera-Damia:2026dbk}. However, let us note some qualitative similarities with SSB in flat space. At finite times, field-strength two-point functions \cref{field-strengt 2-point function any n} are not \ds-invariant, but at late (and early) times they are. So, the states $\ket{\Omega_{n\neq0}}$, although not \ds-invariant themselves, give rise to \ds-invariant two-point functions at very late times. 

Similarly, one can readily understand that two-point functions in the Fourier dual basis, $\ket*{\tilde{\Omega}_\vartheta}$, of zero-particle states \cref{eq:vartheta-vac}, are also \ds-invariant at late times. Moreover, these states are mapped to one another with the action of the symmetry operator $\ex{\ii\alpha \hat{p}}$, as in \cref{eq:Ovartheta-action}. Since all $\ket*{\tilde{\Omega}_\vartheta}$ states are indistinguishable at this level, choosing any one of them would indicate SSB of the $\U(1)$ shift symmetry. Of course, this is not truly the case, as none of these states is a \ds-invariant vacuum, and there is no obvious reason (e.g. cluster decomposition) to pick such a state. 

Furthermore, there is no clear order parameter. In flat space the natural order parameters would be vertex operators, $V_n(x) = \no{\ex{\ii n \phi(x)}}$. A non-vanishing neutral two-point function thereof at large separation would indicate SSB. Here, vertex operator two-point functions decay very rapidly at late times:
\begin{align}
    \mel{\Omega_0}{\hat{V}_{-n}(t,\vec{\theta}) \hat{V}_n\qty(t,\vec{\theta}')}{\Omega_0} \sim \ex{- \alpha\, n^2\, \abs{t}}~, 
\end{align}
for some positive constant $\alpha$, with two-point functions in other $n \neq 0$ zero-particle states behaving similarly. This behaviour can be inferred, for instance, from \cite{Ratra:1985} or \cref{eq:VV-2pt}. We find ourselves in the delicate situation where the would-be order parameter has diluted completely at late times and we lose access to it, but the available observables (e.g. field-strength two-point functions) at late times behave in a manner compatible with symmetry breaking. Barring $(D-1)$-form symmetries, this is likely as close as one can get to SSB in \ds.

\section{The Hilbert space of electromagnetism on \texorpdfstring{\(\ds_3\)}{dS3}}\label{sec:EM-Hilbert}

By the duality between the compact scalar and a $(D-2)$-form discussed in \cref{ssec:axion-many-faces}, the Hilbert space of the massless compact scalar also defines the Hilbert space of the $(D-2)$-form gauge field on $\ds_D$. Some details concerning the equivalence of the two Hilbert spaces are in order for $D=3$, where our scalar is dual to a photon, as already mentioned in \cref{Maxwell equations dS_3}. For discussions on electromagnetism in dS spacetime, see, e.g., \cite{Loparco:2025azm, Gizem, RiosFukelman:2023mgq, Cotescu2021QuantumTO, Loganayagam:2025jmw, Cotaescu:2008hv}, while for discussions on various dualities in dS, see \cite{Hinterbichler:2016fgl, Hinterbichler:2024vyv}.

\noindent  \textbf{Oscillator sector.} We first discuss the oscillator sector of the Hilbert space.
The Maxwell equations in the Lorenz gauge \cref{Maxwell equations dS_3} have a gauge redundancy of the form
\begin{align}
    \delta A_{\mu}(t, \vec{\theta}) = \pd_{\mu}\chi(t, \vec{\theta})~, \qquad\text{with}\qquad \Box\, \chi(t, \vec{\theta})=0~.
\end{align}
As a consequence, the field equations admit pure-gauge solutions of the form%
\footnote{Since in $D=3$ the vector $\vec{m} = (m_{D-2},\cdots, m_1)$ \cref{eq:labels-spherical-harmonics} labelling spherical harmonics contains only one entry, $m_1$, we rename it to $m$ to simplify notation.}
\begin{align}
    A^{(\t{PG})}_{\mu} (t, \vec{\theta})= \pd_{\mu}\chi(t, \vec{\theta}),\qquad \chi \in \spn{\phi_{\ell m},\; 1,\; f_{3}(t)}.
\end{align}
To fully fix the gauge, we work in radiation gauge: $A_{t}=0$, $\nabla^{j}A_{j}=0$ ($j=\theta_1,\theta_2$). The Bunch--Davies (oscillator) vacuum $\ket{0}$ is the state annihilated by all $\hat{a}_{\ell m}$. The fully gauge-fixed potential then is promoted to a quantum field and is expanded in modes as
\begin{align}
    \hat{A}_{j}(t,\vec{\theta}) = \sum_{\ell=1}^{\infty}\sum_{m=-\ell}^\ell \qty(\hat{a}_{\ell m} {A}^{(\ell \, m)}_{j}(t,\vec{\theta}) + \hat{a}_{\ell m}^{\dagger} {A}^{(\ell \, m)}_{j}(t,\vec{\theta})^* ),
\end{align}
where the quantum numbers $\ell,m$ and the creation/annihilation operators are the same as in \cref{quantised oscillator part}.
Here, the modes ${A}^{(\ell m)}_{\mu}(t,\vec{\theta})$ with
\begin{equation}
    A^{(\ell \,m)}_{t}=0~,\qquad \nabla^j A^{(\ell \, m)}_{j}=0~,
\end{equation}
are the ``physical'' spin-1 (positive frequency) Bunch--Davies modes describing the propagating degrees of freedom of the photon \cite{Higuchi:1986wu}. They are obtained through analytic continuation of transverse vector spherical harmonics on $S^3$, and they are given by \cite{Higuchi:1986wu}
\begin{align}\label{eq:A-modes}
    A^{(\ell \, m)}_{j}(t, \vec{\theta}) = C_{\ell}\, (\cosh{t})^{1/2}\; P_{-1/2}^{-(\ell+1/2)}(\ii \, \sinh{t})\; \tilde{Y}^{(\ell \, m)}_{j}( \vec{\theta}), \qquad j=\theta_1, \theta_2~.
\end{align}
Here $\tilde{Y}^{(\ell \, m)}_{j}( \vec{\theta})$  are the transverse vector spherical harmonics on $S^2$
\begin{align}\label{eq:VSH-eigeneq}
    \tilde{\nabla}^i  \tilde{\nabla}_i\; \tilde{Y}^{(\ell \, m)}_{j}(\vec{\theta}) = -\qty\big(\ell(\ell +1) - 1)\, \tilde{Y}^{(\ell \, m)}_{j}( \vec{\theta})~, \qquad \tilde{\nabla}^j\tilde{Y}^{(\ell \, m)}_{j}( \vec{\theta}) =0.
\end{align}
The explicit expression for the normalisation factor $C_{\ell}$ is not needed for our analysis, but can be found in \cite{Higuchi:1986wu}.

The infinitesimal dS transformations of the modes are encoded by the Lie derivative with respect to Killing vectors 
\begin{align}
    \lie_{\xi}A_{\mu} = \rds\, \qty(  \xi^\rho\,\nabla_{\rho}\,A_{\mu}  + A_{\rho}\nabla_{\mu}\,\xi^{\rho} )~.
\end{align}
The set of physical modes furnishes the same principal-series $\SO(3,1)$ UIR ($\Delta=s=1$) as the one furnished by the massless scalar modes  (cf. \cref{Subsec:UIRs-BD-modes}). The  time-independent \ds-invariant inner product, with respect to which the modes form the UIR, is the natural generalisation of the Klein--Gordon inner product. It is given by \cite{Higuchi:1986wu}:
\begin{align}
    \ip{A^{(1)}}{{A^{(2)}}}
    = \rds\,\ii\cosh^{2}t\int_{\S^{2}}\dvol_{2}\,
    \qty[
        \qty(A^{(1)\mu})^*\,\nabla_t\, A^{(2)}_{\mu}
        -\qty(\nabla_t\, A^{(1)\mu})^*\,A^{(2)}_{\mu}
    ]~.
\end{align}
The photon Bunch--Davies modes are orthonormal under this inner product,
\begin{align}
    \ip{A^{(\ell m)}}{A^{(\ell' m')}}
    = \delta_{\ell\ell'}\delta_{m\,m'}~, \quad 
     \ip{A^{(\ell m)*}}{A^{(\ell'm')*}}
    = -\delta_{\ell\ell'}\delta_{m\,m'}~, \quad
    \ip{A^{(\ell m)}}{A^{(\ell'm')*}}
    = 0~.
\end{align}
Pure-gauge modes are orthogonal to all modes, including themselves,
\begin{align}
    \ip{A^{(\t{PG})}}{A^{(\ell m)}}
    =
    \ip{A^{(\t{PG})}}{A^{(\ell m)*}} = 0~, \qquad
    \ip{A^{(\t{PG})}}{A^{(\t{PG}')}} = 0~,
\end{align}
where $A^{(\t{PG})}_{\mu}$ and $A^{(\t{PG}')}_{\nu}$ are any two pure-gauge modes.
Furthermore, pure-gauge modes transform only among themselves under dS transformations. In the vector space of physical modes, they are identified with zero due to the equivalence relation: $A^{(\ell \, m)}_{\mu} \sim A^{(\ell \, m)}_{\mu} + A_{\mu}^{(\t{PG})}$.


\paragraph{Monopole sector.} To reproduce the duality of electromagnetism with the compact massless scalar theory at the quantum level, the corresponding Hilbert spaces should come out isomorphic. How the oscillator sector matches was made clear above. Let us now see how the zero-mode sector of the massless compact scalar manifests itself {as a monopole sector} from the electromagnetic viewpoint.

Because global $\ds_3$ contains a non-trivial 2-cycle --- its spatial slice, $\S^2$ --- the field strength $\hat{F}_{\mu \nu}$ is not simply expressed as the exterior derivative of a (globally well-defined) gauge potential. In particular, since $\hat{F}_{\mu \nu}$ is a closed two-form on global $\ds_3$ we have:
\begin{align}
    \hat{F}_{\mu \nu} =  \hat{\cF}_{\mu \nu} + \hat{F}^{(\t{osc})}_{\mu \nu} = \hat{\cF}_{\mu \nu} + 2 \pd_{[\mu}  \hat{A}_{\nu]}~.
\end{align}
Here $\hat{\cF}_{ \mu \nu}$ is the harmonic piece of the field strength, satisfying
\begin{equation}
    \pd_{[\mu} \hat{\cF}_{\rho\sigma]} = 0 \qq{and} \nabla^{\mu} \hat{\cF}_{\mu \nu} = 0~,
\end{equation}
off-shell. The operator $\hat{A}_{\mu}$ is the standard globally well-defined oscillator gauge-potential one-form. 

The duality \cref{eq:duality-D-dim} specialised to $D=3$ and written in Lorentzian signature, reads:
\begin{align} 
\hat{F}_{\mu\nu} = 2\pi f^2\, \epsilon_{\mu\nu\rho}\, \partial^\rho \hat{\phi}~.
\end{align}
The oscillator piece, $\hat{\varphi}$, of the scalar field is directly mapped to $\hat{F}^{(\t{osc})}_{\mu\nu}$. This is encoded, for instance in \cref{Maxwell equations dS_3}. The (time-dependent) zero-mode of the scalar sees the harmonic part of the electromagnetic field strength:
\begin{align} 
\hat{\cF}_{\mu\nu} = 2\pi f^2\, \epsilon_{\mu\nu\rho}\, \partial^\rho \hat{\phi}_{0}(t)~.
\end{align}
Utilising this, it follows that the harmonic field strength has vanishing electric component, $\hat{\cF}_{ti}=0$, while its spatial components are:
\begin{align}
    \hat{\cF}_{ij}   &= 
    -\frac{2\pi\, \hat{p}}{\rds^2\,\Vol_2} \frac{\epsilon_{tij}}{\cosh^2{t}} = -\frac{2\pi\,\hat{p}}{\rds^2\,\Vol_2} \tilde{\epsilon}_{ij}~.
\end{align}
This corresponds to the magnetic field of a monopole:
\begin{align}
    \hat{\cB} \coloneqq \frac{1}{2}\epsilon_{t\mu\nu}\hat{\cF}^{\mu\nu} = -\frac{2\pi\,\hat{p}}{\rds^2\,\Vol_2\,\cosh^2{t}} = -\frac{\hat{p}}{2 \rds^2\cosh^2{t}}~.
\end{align}
In fact, expressing $\hat{F}_{\mu\nu}$ in Maxwell's equations
\begin{align}
    \nabla^{\mu}\hat{F}_{\mu \nu}=0~,\qquad \nabla^{\mu}\qty\big(\star{\hat{F}})_\mu=0~,
\end{align}
as $\hat{F}_{\mu \nu} = \hat{\cF}_{\mu \nu}+\hat{F}^{(\t{osc})}_{\mu \nu}$, we find
\begin{align}
\nabla^{\mu}\qty(\star{\hat{F}}^{(\t{osc})})_\mu= \frac{\partial \hat{\cB}}{\partial t}~.
\end{align}
Hence  $\pdv*{\hat{\cB}}{t}$ is a magnetic current sourcing Maxwell's equations for the oscillator part.

From the above discussion, it follows that the operator $\hat{p}$ is exactly the monopole charge. Its eigenvalues are quantised due to Dirac quantisation.
Its conjugate, $\hat{q}$, or rather the well-defined operator $\ex{\ii n \hat{q}}$, cannot be defined locally in terms of the electric and magnetic fields. From the electromagnetic point-of-view, $\ex{\ii n \hat{q}}$ is a disorder operator \cite{Kadanoff:1970kz,Fradkin:2016ksx}, defined by removing a point from spacetime and fixing the magnetic flux around a small sphere surrounding it to be $2\pi n$. While its description is complicated in this frame, it crucially provides the conjugate variable to $\hat{p}$. In total, this conjugate pair provides the missing $L^2(\S^1)$ piece of the Hilbert space. We will not repeat the quantisation here, as the details are similar to \cref{sec:canonical}.

Before concluding this section, we note that the harmonic form $\hat{\cF}_{\mu \nu}$ can be expressed in terms of a \emph{non-globally well-defined} gauge potential $\hat{\cF}_{\mu \nu} = 2 \partial_{[\mu} \hat{\cA}_{\nu]}$ à la Dirac \cite{Dirac:1931kp}. The gauge field has components:
\begin{align}
\hat{\cA}_{t}=0~,\qquad 
\hat{\cA}_{\theta_2}=0~,\qquad 
\hat{\cA}_{\theta_1}=-\frac{2\pi\hat{p}}{\Vol_{2}}\,(1+\cos{\theta_2})~.
\end{align}
This potential satisfies $\nabla^{\mu}\hat{\cA}_{\mu}=0$. It can be readily checked that this gauge potential gives the correct field strength. However, it is not well-defined at the north pole of the spatial $\S^2$, as can be easily checked.%
\footnote{Had this been a well-defined potential, its spatial part would have been expressed as  $\cA_i = \tilde{\epsilon}_{ij} \widetilde{\nabla}^j \psi(\theta_1,\theta_2)$, where $i,j\in\set{\theta_1,\theta_2}$ and $\psi(\theta_1,\theta_2)$ is a smooth scalar function on $S^2$. In our case, the scalar function turns out to have the form $\psi \propto \log(\sin(\theta_2/2))$, which is singular at the north pole.}
Of course, there is also a construction in the spirit of Wu and Yang \cite{Wu:1975es}, involving two well-defined gauge fields, each defined on a portion of the spatial sphere and differing on overlapping regions by a gauge transformation. 

\section{Euclidean sphere path integrals}\label{sec:sphere}

Having analysed the physics of compact scalars (axions) in de~Sitter spacetime from a Lorentzian perspective, there is a clear takeaway message. Besides the usual vacuum, there is a discretuum of zero-particle states, $\ket{\Omega_n}$, labelled by all allowed charged configurations on the spatial $(D-1)$-sphere. Above we have linked that to the topology of the spatial slice, having a non-trivial $(D-1)$-cycle.

This interpretation presents a potential puzzle for readers accustomed to Euclidean path integral methods for QFT in \(\ds\). The reason is that global \(\ds_D\) Wick-rotates to the Euclidean $D$-sphere, \(\S^D\), which contains no $(D-1)$-cycles.%
\footnote{Apart from the trivial $D=1$ case, in which case the Lorentzian geometry is just the real line.}
The sphere partition function can thus not capture the topological contributions we uncovered earlier. The aim of this section is to resolve this apparent puzzle and provide a Euclidean calculation which does capture these contributions. In other words, we will address the following question: is there a local Euclidean path integral that captures the entire Hilbert space?

In what follows, we adopt units where $\rds=1$, i.e. we work with the unit $S^{D}$. 


\subsection{Sphere path integral and Harish-Chandra characters}\label{Subsec: path integral and characters}
The 1-loop sphere path integral of fields placed on $S^D$ can be expressed in terms of Harish-Chandra characters of  a boost group element of the \emph{Lorentzian} de~Sitter group $\SO(D,1)$, as shown recently in \cite{Anninos:2020hfj}.%
\footnote{See also \cite{Anninos:2025mje, Anninos:2026hia, Law:2026tuk, David:2021wrw, Ball:2024hqe, Anninos:2023exn}.}
Here, we focus on the sphere path integral of the compact massless minimally coupled scalar on $S^D$,
\begin{align}
    \parti_{\S^D} = \int \DD{\phi}\;  \ex{-S_\t{E}[\phi]}~, \qquad
    S_\t{E}[\phi]
    = \frac{\kappa_D}{2}\int \dd[D]{x}\sqrt{g}\;g^{\mu\nu}\pd_\mu\phi\,\pd_\nu\phi~,
\end{align}
which corresponds to the 0-form case studied in \cite{David:2021wrw}. The path integral has the standard expression in terms of a functional determinant
\begin{align}
   \parti_{\S^D} = \frac{\sqrt{2\pi \Vol_{D}\, \kappa_D}}{\sqrt{\det'(-\nabla^{\mu} \nabla_{\mu})}}~,
\end{align}
where the prime in $\det'$ denotes omission of the vanishing eigenvalue of the Laplace-Beltrami operator $\nabla^\mu \nabla_\mu$ associated with the zero mode of the scalar. Since the scalar is compact, path integration over the zero mode can be carried out explicitly and gives rise to the factor in the numerator.

The sphere path integral of the compact massless scalar, $ \parti_{\S^D}$, can be expressed in terms of a  ``naive'' (in the sense of \cite{Anninos:2020hfj}) bulk character $\hat{\chi}^{\t{bulk}}(t)$ \cite{David:2021wrw}:
\begin{align}
    \log{\parti_{\S^D}} = \frac{1}{2} \log(2 \pi \Vol_{D}\, \kappa_D)+\log{\parti_{S^{D}}'}~,
\end{align}
where
\begin{align}
  \log{\parti_{S^{D}}'} = \int_0^{\infty} \frac{\dd{t}}{2t} \frac{1+\ex{-t}}{1-\ex{-t}}\; \hat{\chi}^{\t{bulk}}(t)~, \qquad \hat{\chi}^{\t{bulk}}(t)=\frac{\ex{-(D-1)t}+1}{\qty(1-\ex{-t})^{D-1}}~.
\end{align}
Following \cite{Anninos:2020hfj}, we would like to extract the actual  $\SO(D,1)$ Harish-Chandra character from the naive character $\hat{\chi}^{\t{bulk}}(t)$ by applying the ``flipping procedure''. Note that the ``flipped character'', i.e. the actual Harish-Chandra character associated with UIRs of $\SO(D,1)$, was not identified in \cite{David:2021wrw} in the case of 0-forms, but it was discussed for higher forms. Applying the flipping procedure \cite{Anninos:2020hfj}, we find  that the naive character is expressed as
\begin{align}
    \hat{\chi}^{\t{bulk}}(t) =  {\chi}^{\t{bulk}}(t)+1,
\end{align}
where ${\chi}^{\t{bulk}}(t)$ is the $\SO(D,1)$ Harish-Chandra character of the (oscillator) single-particle UIR%
\footnote{See, e.g., \cite{Penedones:2023uqc}, as well as \cref{Subsec:UIRs-BD-modes} for the identification of the UIR for all $D$.}
\begin{align}
  {\chi}^{\t{bulk}}(t) =  \frac{\ex{-(D-1)t}+1}{\qty(1-\ex{-t})^{D-1}}-1~.
\end{align}
To conclude, the sphere path integral takes the form
\begin{align}\label{eq:character-represenation}
  \log{\parti_{S^{D}}} 
  &=
  \int_0^{\infty} \frac{\dd{t}}{2t} \frac{1+\ex{-t}}{1-\ex{-t}}\; {\chi}^{\t{bulk}}(t) \nonumber\\ 
  &\phantom{=~}+\int_0^{\infty} \frac{\dd{t}}{2t} \frac{1+\ex{-t}}{1-\ex{-t}}\; \times 1 +\frac{1}{2} \log(2 \pi \, \Vol_{D}\, \kappa_D).
  \end{align}
  This is the character integral representation of the sphere partition function for the compact massless scalar.
  
\subsection{The Euclidean vacuum}

On general grounds, Euclidean correlators on a \(D\)-dimensional sphere can be viewed upon analytic continuation as Lorentzian correlators in a \(\ds_D\)-invariant state, that is:
\begin{align}
	\ev{\cO_1(\tau_1,x_1)\cO_2(\tau_2,x_2)\cdots}_{\S^d} = \mel{\t{E}}{\cO_1(t_1,x_1)\cO_2(t_2,x_2)\cdots}{\t{E}}~,
\end{align}
where \(t_\ell\) are related by analytic continuation to \(\tau_\ell\) as: \(\tau_\ell = \pi/2-\ii\, t_\ell\). Crucially, in the above, the state \(\ket{\t{E}}\) is invariant under all de~Sitter isometries. To see this, it suffices to note that the \(\S^D\) correlators satisfy \(\SO(D+1)\) Ward identities, which, upon analytic continuation, become \(\SO(D,1)\) Ward identities. Hence, \(\ket{\t{E}}\) must also be \(\ds\)-invariant. We will refer to the state $\ket{E}$ as ``Euclidean vacuum'', but it is also known as ``Bunch--Davies state'', or in more quantum-gravity-oriented literature ``Hartle--Hawking state''.

We note here that, conventionally, both the correlator and the Euclidean vacuum are chosen unnormalised. In particular, the norm of the Euclidean vacuum is precisely the empty sphere path integral:
\begin{align}
	\braket{\t{E}} = \ev{1}_{\S^D} = \parti_{\S^D}
\end{align}
Pictorially, the Euclidean vacuum can be viewed as the state prepared by the empty path integral on a hemisphere \cite{Tolley:2001gg,Miller:2025jbz}:
\begin{align}
	\ket{\t{E}} =
	\begin{gathered}
		\begin{tikzpicture}[scale=0.8]
			\draw  (0,4) ++ (0:2 and 1) arc (0:180:2 and 1);
			\draw  (0,4) ++ (180:2 and 1) arc (180:360:2 and 1);
			\draw  (0,4) ++ (180:2 and 1) arc (180:360:2 and 2);
		\end{tikzpicture}
	\end{gathered}~.
\end{align}
This picture enables a further immediate conclusion: \(\ket{\t{E}}\) can also not carry charge under any (internal) global symmetry of the underlying QFT. To connect with the discussion above, in \cref{sec:canonical} we found a unique \(\ds\)-invariant state. We may, then, identify:
\begin{align}
    \frac{1}{\sqrt{\parti_{\S^D}}}\;\ket{\t{E}} = \ket{\Omega_0} = \ket{0}_\t{QM} \otimes \ket{0}~,
\end{align}   
in the notation of \cref{sec:canonical}. In other words, \(\ket{\t{E}}\) is, up to normalisation, the state with zero charge under the \(\U(1)\) shift symmetry of the scalar, and all the oscillators at zero occupation number. As a further check, this state is annihilated by all the lowering Bunch--Davies modes, $\hat{a}_{\ell\vec{m}}$, \cite{Miller:2025jbz}. It is then clear that the bare sphere partition function can only capture states in the Fock space, $\cF_\t{osc}$, above \(\ket{\Omega_0}\)\footnote{See \cite{Tolley:2001gg} for a related discussion in the case of the non-compact scalar.}. This is consistent with the character representation of the sphere partition function \cref{eq:character-represenation}.

\subsection{Charged zero-particle states}

To describe states with non-zero shift charge one needs to insert a charged operator on the hemisphere. In two dimensions, the answer would be immediate by the state-operator correspondence and the Weyl equivalence of the hemisphere and the disc \cite{DiFrancesco:1997nk}. In higher dimensions, the question of state-operator correspondence was analysed in \cite{Vitouladitis:2025zoy},\footnote{See also \cite{Hofman:2024oze,Arbalestrier:2026lna} for generalisations to extended operators.} where it was shown that states of charge \(n\) are prepared by a path integral on a ball with the insertion of a vertex operator%
\footnote{Note that in the path integral, normal ordering $\no{\cdots}$ denotes renormalisation of the operator to remove coincident-point divergences.}
\begin{align}
	V_n(x) = \no{\ex{\ii n \phi(x)}}~,
\end{align}
on the south pole. As far as the charge of the state is concerned, the analysis of \cite{Vitouladitis:2025zoy} goes through unchanged. It follows that charged zero-particle states are identified with a hemisphere path integral with an insertion on the south pole, as
\begin{align}\label{eq:n-state-PI}
	\ket{\Omega_n} =
	\frac{1}{\sqrt{\parti_n}}
    \begin{gathered}
		\begin{tikzpicture}[scale=0.8]
			\draw  (0,4) ++ (0:2 and 1) arc (0:180:2 and 1);
			\draw  (0,4) ++ (180:2 and 1) arc (180:360:2 and 1);
			\draw  (0,4) ++ (180:2 and 1) arc (180:360:2 and 2);
			\begin{scope}[even odd rule]
				\clip (0,4) ++ (180:2 and 1) arc (180:360:2 and 2);
				\draw [fill=black] (0,2) circle (.1);
			\end{scope}
			\draw (0.1,2) node [below] {$V_n$};
		\end{tikzpicture}
	\end{gathered}~.
\end{align}
The normalisation factor, \(\parti_n\), is given by the sphere two-point function of \(V_n\) and its Hermitian conjugate inserted at the two poles of the sphere:
\begin{align}\label{eq:VV-2pt-fn}
        \cZ_n = \ev{V_n^\dagger(\t{N})\, V_n(\t{S})}_{\S^D}~.
\end{align}

The only difference relative to the case studied in \cite{Vitouladitis:2025zoy} lies in the set of annihilation operators that annihilate the states. In particular, the state in \cref{eq:n-state-PI} is squeezed relative to its analogue in \cite{Vitouladitis:2025zoy}. In the present case, \(\ket{\Omega_n}\) is annihilated precisely by the Bunch--Davies annihilation operators, as can be inferred by a variant of the arguments in \cite{Miller:2025jbz}. More explicitly, let us use the Lorentzian operator algebra to show this. First, the zero-mode part of \(\hat{\phi}\) commutes with all \(\hat a_{\ell\vec{m}}\), \(\ell \geq 1\). For the non-zero modes, the oscillator expansion \cref{quantised oscillator part} gives
\begin{align}
    \comm{\hat{a}_{\ell\vec{m}}}{\hat{\varphi}(x)}
    =
    \frac{1}{\sqrt{\kappa_D}}\,
    \phi_{\ell\vec{m}}^*(x)~.
\end{align}
For the normal-ordered vertex operator this implies, at Lorentzian points \(x\),
\begin{align}\label{eq:comm-vertex-osc}
    \comm{\hat a_{\ell\vec{m}}}{\hat{V}_n(x)}
    =
    \frac{\ii n}{\sqrt{\kappa_D}}\,
    \phi_{\ell\vec{m}}^*(x)\,\hat{V}_n(x)~.
\end{align}
The insertion in \cref{eq:n-state-PI} is obtained by analytically continuing the vertex operator to the south pole of the sphere. We must therefore analytically continue \cref{eq:comm-vertex-osc}, and finally evaluate it, at the south pole of the hemi-$S^D$. From \cref{eq:bd-modes-dsd} and upon complex-conjugation, we can read off the continuation of $\phi_{\ell\vec{m}}^*$ as:
\begin{align}\label{eq:bd-modes-sphere-minus-south}
    \phi_{\ell\vec{m}}^{*,\t{E}}(\tau,\vec{\theta})
    =
    \mathcal{N}_{\ell,D}\,
    \qty(\sin\tau)^{-\frac{D-2}{2}}\,
    P_{\frac{D-2}{2}}^{-\qty(\ell+\frac{D-2}{2})}(-\cos\tau)\,
    Y_{\ell\vec{m}}^*(\vec{\theta})~, \quad 0\leq \tau\leq \pi~.
\end{align}
Further using the asymptotic behaviour of the associated Legendre polynomials \cite[§14.8]{DLMF}, 
\begin{align}
    P_\nu^{-\mu}(\cos u)
    \sim
    \frac{1}{\Gamma(\mu+1)}
    \qty(\frac{u}{2})^\mu~,
    \qquad
    u\to 0~,
\end{align}
we find that the coefficient in \cref{eq:comm-vertex-osc} vanishes at the south pole%
\footnote{Note that the modes $\phi_{\ell\vec{m}}^{*,\t{E}}$ have singularities at the north pole. The situation is reversed for the analytic continuation of the positive-frequency modes. (Note that $P_\nu^\mu(-x)\kern-2pt\not\kern1pt\propto P_\nu^\mu(x)$ when $\nu+\mu$ is a negative integer.)
Hence the empty upper-hemisphere path-integral prepares the bra state $\bra{\t{E}}\propto\bra{\Omega_0}$, and insertions of $V_n^\dagger$ prepare $\bra{\Omega_n}$.}
\begin{align}
    \phi_{\ell\vec{m}}^{*,\t{E}}(\tau,\vec{\theta})
    \sim
    \frac{\mathcal{N}_{\ell,D}}
    {2^{\ell+\frac{D-2}{2}}\Gamma\qty(\ell+\frac{D}{2})}\,
    \qty(\pi-\tau)^\ell\,Y_{\ell\vec{m}}^*(\vec{\theta})~,
    \qquad
    \tau\to\pi~.
\end{align}
Hence \(\comm{a_{\ell\vec{m}}}{V_n(\t{S})}=0\).%
\footnote{While we continue to write the commutator, this expression should rather be interpreted as an operator equation, valid for path-integral insertions. See also \cite{Miller:2025jbz,Vitouladitis:2025zoy,Arbalestrier:2026lna}.}
Using also \(\hat a_{\ell\vec{m}}\ket{\t{E}}=0\), and that $\ket{\Omega_n} \sim V_n(\t{S}) \ket{\t{E}}$, it follows that
\begin{align}
    \hat a_{\ell\vec{m}}\ket{\Omega_n}
    =
    \frac{1}{\sqrt{\parti_n}}\,
    \hat a_{\ell\vec{m}}V_n(\t{S})\ket{\t{E}}
    =0~.
\end{align}
Acting instead with \emph{creation} operators, one may then reach any state in the Fock space built on top of \(\ket{\Omega_n}\).

Just as the sphere partition function encodes the usual Hilbert space built on the Bunch--Davies vacuum, the two-point function \(\parti_n\), given in \cref{eq:VV-2pt-fn}, encodes the Hilbert space of charge-\(n\) states. This two-point function was studied in detail in \cite{Chakraborty:2023eoq,Chakraborty:2025mhh} and takes the form%
\footnote{Our normalisation differs slightly from that of \cite{Chakraborty:2023eoq}. After removing the coincident-point divergence, the vertex operators in \cite{Chakraborty:2023eoq} are further multiplied by a finite factor, chosen so that their two-point function has a prescribed behaviour after analytic continuation. We do not include this additional rescaling. Explicitly,
\begin{equation*}
	V_n^{\text{there}}(x)
	=
	\exp(
	\frac{n^2\; \qty(\sfH_{D-1}+\sfH_{D-2}-\sfH_{(D-2)/2})}{2(D-1)\Vol_{D}\,\kappa_D}
	)
	V_n^{\text{here}}(x).
\end{equation*}
}
\begin{align}\label{eq:VV-2pt}
    \ev{V_n^\dagger(x)\;V_n(y)}_{\S^D}
    =
    \parti_n
    \exp(
        \frac{n^2}{D\,\kappa_D\,\Vol_{D}}\,
            {}_3F_2\qty[\mqty{1,\ 1,\ D \\ 2, (D+2)/2}\suchthat\frac{1+\xi}{2}]\; (1+\xi)
        )~.
\end{align}
Here \(\xi=\xi(x,y)\) is the \(\SO(D+1)\)-invariant distance between $x$ and $y$, defined in \cref{eq:sphere-invariant-xi}. Note also that $\ev*{V_m^\dagger(x) V_n(y)}=0$ if $m\neq n$. The prefactor $\parti_n$ encodes the two-point function at antipodal points, \cref{eq:VV-2pt-fn}, for which $\xi=-1$. It takes the form:
\begin{align}\label{eq:Zn-in-terms-of-q}
    \parti_n = \sfq^{n^2}\;\parti_{\S^D}~, \qq{with} \sfq\coloneqq \exp(-\frac{\sfH_{D-1}}{(D-1)\,\kappa_D\,\Vol_{D}})~,
\end{align}
where $\sfH_{D-1}$ is the $(D-1)$-th harmonic number.

\subsection{A path integral for the full Hilbert space}

So far we have understood how each individual \(\parti_n\) captures the zero-particle state \(\ket{\Omega_n}\) in the sense of eq. \cref{eq:n-state-PI}. Also, from eq. \cref{eq:Zn-in-terms-of-q}, we understand that the zero-mode contribution of \(\parti_n\)  appears through the factor \(\sfq^{n^2}\), while the normalisation \(\parti_{\S^D}\) captures the oscillator contribution (through the Harish-Chandra characters of the associated $SO(D,1)$ UIR, see \cref{Subsec: path integral and characters}). But the question with which we began this section still remains: is there a local Euclidean path integral that captures the entire Hilbert space, in the same sense as the standard sphere path integral captures the oscillator sector? The answer is yes. Consider the state
\begin{equation}
	\ket{\cO_\vartheta} \coloneqq 
    \begin{gathered}
		\begin{tikzpicture}[scale=0.8]
			\draw  (0,4) ++ (0:2 and 1) arc (0:180:2 and 1);
			\draw  (0,4) ++ (180:2 and 1) arc (180:360:2 and 1);
			\draw  (0,4) ++ (180:2 and 1) arc (180:360:2 and 2);
			\begin{scope}[even odd rule]
				\clip (0,4) ++ (180:2 and 1) arc (180:360:2 and 2);
				\draw [fill=black] (0,2) circle (.1);
			\end{scope}
			\draw (0.1,2) node [below] {$\cO_\vartheta$};
		\end{tikzpicture}
	\end{gathered}~, \qq{with} \cO_\vartheta(x) = \sum_{n\in\Z} \ex{-\ii\vartheta n}\,V_n(x)~,
\end{equation}
where $\cO_\vartheta$ is inserted at the south pole. These are an unnormalised version of the states \cref{eq:vartheta-vac}.
Their norm, i.e. the path integral with insertions of \(\cO_\vartheta\) at antipodal points,%
\footnote{Note that \(\cO_\vartheta\) is Hermitian: \(\cO_\vartheta^\dagger = \cO_\vartheta\).}
is given by
\begin{align}\label{eq:ultra-grand-canonical-parti}
	\widehat{\parti}_{\S^D} \coloneqq \ev{\vphantom{V_n^\dagger}\cO_\vartheta(\t{N})\,\cO_\vartheta(\t{S})}_{\S^d} = \parti_\t{QM}\;\parti_{\S^D}~.
\end{align}
Here
\begin{align}\label{eq:parti-QM-PI}
  \parti_\t{QM}=  \sum_{n \in \mathbb{Z}} \sfq^{n^2} = \theta\qty(\sfq)~,
\end{align}
where $\theta$ denotes the Jacobi theta function, and $\sfq$ is as in \cref{eq:Zn-in-terms-of-q}. In this way, the path integral $\widehat{\parti}_{S^D}$ factorises into two parts. One factor is the standard sphere partition function, $\parti_{\S^D}$, which captures the oscillators, but ignores the topological sectors. The other factor, $\parti_\t{QM}$, captures exactly the zero-mode/topological sectors.

In fact, \(\parti_\t{QM}\) is exactly the quantum-mechanical partition function of a free particle on a circle, i.e. a quantum rotor. This becomes evident by re-expressing it as 
\begin{align}
	\parti_\t{QM} = \tr\ex{-\beta\hat{H}}~, \qq{with} \hat{H} = \frac{\hat{p}^2}{2\,I}~,
\end{align}
with the trace taken over ${L^2(S^1)}$. This agrees with \cref{eq:parti-QM-PI} upon identifying $\sfq = \ex{-\beta/(2I)}$. In the above, $I$ is the moment of inertia of the quantum rotor and $\beta$ is the inverse temperature at whcih the system is placed. In this sense, \cref{eq:ultra-grand-canonical-parti} is the path integral incarnation of the canonical result \(\cH=L^2(\S^1)\otimes \cF_\t{osc}\) \cref{eq:compact-scalar-hilbert-space-dsd}. 

A remarkable consequence of \cref{eq:ultra-grand-canonical-parti} is the emergence of a thermal circle directly from the sphere path integral. In particular, comparing the quantum-mechanical answer with \cref{eq:Zn-in-terms-of-q} identifies the ratio $\beta/(2\,I)$ as 
\begin{align}
    \frac{\beta}{2\,I} = \frac{\sfH_{D-1}}{(D-1)\kappa_D \Vol_{D}}~.
\end{align} 
Moreover, matching this Hamiltonian to the zero-mode Hamiltonian \cref{eq:particle-ring-hamiltonian}, at $t=0$ --- where one could consider initiating a Lorentzian evolution of the state prepared --- fixes the moment of inertia as:
\begin{align}
    I = \rds\, \kappa_D \Vol_{D-1}~.
\end{align}
It follows that the particle is at effective inverse temperature:
\begin{align}
    \beta = \frac{2\,\sfH_{D-1}\Vol_{D-1}}{(D-1)\Vol_{D}}\,\rds = \frac{\Gamma\qty(\frac{D-1}{2}) \qty(\psi(D)+\gamma_\t{E})}{\sqrt{\pi}\, \Gamma\qty(\frac{D}{2})}\,\rds~,
\end{align}
where $\psi$ is the digamma function and $\gamma_\t{E}$ is the Euler--Mascheroni constant. For $D=3$ and $D=4$ the corresponding inverse temperatures are:
\begin{align}
    \beta_{D=3} = \frac{3}{\pi}\,\rds~, \qq{and} \beta_{D=4} = \frac{11}{12}\,\rds~.
\end{align}

Traditionally, the sphere path integral of a given field is believed to capture thermal properties of the Lorentzian QFT in the static patch of $\ds_D$, since the static patch can be Wick-rotated to $S^{D}$, see e.g. \cite{Grewal:2024jes, Anninos:2020hfj}. Moreover, the Harish-Chandra characters that appear at one loop have been shown to encode information about the spectrum of quasinormal modes, which are characteristic traits of the static patch \cite{Anninos:2020hfj}. However, global $\ds_D$ also Wick-rotates to $S^{D}$, while Harish-Chandra characters are associated with the single-particle UIRs of $\SO(D,1)$ which are better understood in global Lorentzian de~Sitter. One can thus expect that the sphere path integral contains information about Lorentzian QFTs in global de~Sitter as well. Interestingly, in the previous paragraphs, we saw that the sphere path integral, decorated with insertions of local operators, also encodes thermal properties (through the apparent temperature of the rotor partition function) even if obtained as the continuation of global de~Sitter.  At present we do not have a deeper understanding of this curious phenomenon. 

A further observation is that the decorated partition function \cref{eq:ultra-grand-canonical-parti} naturally bears the interpretation of a character. To see this, recall the situation in 2d CFT, where the torus partition function of the compact boson carries integrable representations of an Abelian Kac--Moody algebra, and Jacobi theta functions arise as affine characters \cite{DiFrancesco:1997nk}. Recent work suggests that related infinite-dimensional current algebras continue to organise the spectrum of the compact scalar in \(d\neq 2\) as well \cite{Vitouladitis:2025zoy}, and thermal partition functions appear as affine characters of the corresponding representations \cite{Fliss:2023uiv}. From that perspective,  it is tempting to interpret the factor \(\parti_\t{QM}=\theta(\sfq)\)  as some kind of character (of some algebra, yet to be made precise) associated with the zero-mode Hilbert space. 

\section{Outlook}

In this work we studied axions on a fixed de~Sitter spacetime. While there are many questions we left unanswered, the most interesting one is arguably to let spacetime itself fluctuate. Coupled to gravity, axions are known to source new saddles, such as Euclidean axion-wormholes \cite{Giddings:1987cg}. In fact, such wormholes are not semiclassical in the axion frame \cite{Witten:2026twr}, necessitating a quantum account of the axion dynamics. Recent work suggests that axions also support new saddles in spacetimes with positive cosmological constant \cite{Aguilar-Gutierrez:2023ril, FRW-braket}. These saddles have rich topology, making them natural testbeds for revisiting the questions we answered here on \(\ds_D\) and \(\S^D\).  

As a first approximation to the gravitational path integral with \(\Lambda>0\), one can study one-loop contributions around all such saddles, see \cite{Ivo:2026ijv} for a recent account of some. It seems likely that, although the empty sphere path integral is blind to the charged sectors, 
subleading saddles are sensitive to them (whenever such charged sectors are allowed by the saddle topology). Indeed, the ``kettleball'' saddles of \cite{Aguilar-Gutierrez:2023ril} have topology \(\S^1\times\S^{D-1}\), and therefore probe the shift-symmetry sectors. Interestingly, they also have access to the dual \((D-2)\)-form symmetry associated with the conservation of higher-dimensional vortices, an effect present neither in Lorentzian \(\ds_D\) nor in the leading Euclidean sphere saddle. This programme would give a concrete handle on the gravitational path integral in spacetimes with positive cosmological constant. A related approach appears in \cite{Anninos:2025ltd}, in which the dynamics around new instantons made available in Einstein--Maxwell theory \cite{Brill:1992ce} is investigated. The interplay between additional fields and couplings, and the emergence of new Euclidean saddles%
\footnote{Whether such saddles admit a Lorentzian interpretation is an interesting question in its own right.}
deserves further investigation.

\paragraph{Acknowledgements.} It is a great pleasure to thank Dionysios Anninos for  useful discussions and comments. We would also like to thank Charis Anastopoulos, Adrien Arbalestrier, Riccardo Argurio, Giorgos Batzios, Jackson Fliss, Atsushi Higuchi, Elise Paznokas, Eduardo García-Valdecasas, and Gonzalo Villa for useful discussions. We are especially grateful to  Nicole Righi for useful discussions and for guidance regarding the code of conduct during the workshop \href{https://indico.sns.it/event/129/}{\emph{Expanding thoughts on de~Sitter}}. We also thank the organisers of the workshop, Nicole Righi, Alan Rios Fukelman, Carlos Duaso Pueyo, and Paolo Benincasa, as well as its participants for stimulating discussions and encouragement that played an important role in finalising this paper.

The work of V.A.L. is supported by the ULYSSE Incentive Grant for Mobility in Scientific Research [MISU] F.6003.24, F.R.S.-FNRS, Belgium. The work of S.V. is supported by a Marina Solvay fellowship and by the Fonds de la Recherche Scientifique (FNRS) under grant no. 4.4503.15.
\appendix 

\section{Details on single-particle UIRs in the oscillator sector}
\label{App: single particle UIRs in Hilbert space}

In this appendix, we use the representation-theoretic properties of the Bunch--Davies modes given in eqs. \cref{dS-d transf of normalised modes ell>=2,eq:coeff-A-lm,dS-d transf of nrmlsd modes ell=1,eq:tildeA1m,eq:c0} to show that the single-particle Hilbert space of the massless compact scalar field furnishes  the same $\SO(D,1)$ UIR as the one formed by the Bunch--Davies modes (see \cref{Subsec:UIRs-BD-modes}).

First, we assign the dS transformation $\lie_{\xi} \hat{\varphi}$ of the oscillator part of the field to the creation and annihilation operators in the mode expansion \cref{quantised oscillator part}, as
\begin{align}
\delta_{\xi} \hat{\varphi}(t,\vec{\theta})
\coloneqq \lie_{\xi}  \hat{\varphi}(t, \vec{\theta})
&= \frac{1}{\sqrt{\kappa_D}} \sum_{\ell=1} \sum_{\vec{m}} \hat{a}_{\ell \vec{m}} \lie_{\xi} \phi_{\ell \vec{m}}(t, \vec{\theta}) +\hat{a}^{\dagger}_{\ell \vec{m}} \lie_{\xi} \phi^{*}_{\ell \vec{m}}(t, \vec{\theta}) \nonumber \\ 
&= \frac{1}{\sqrt{\kappa_D}} \sum_{\ell =1} \sum_{\vec{m}} \delta_{\xi}\hat{a}_{\ell \vec{m}}\,\phi_{\ell \vec{m}}(t, \vec{\theta}) +\delta_{\xi}\hat{a}^{\dagger}_{\ell m}\,\phi^{*}_{\ell \vec{m}}(t, \vec{\theta})~,
\end{align}
where $\xi^{\mu}$ is any dS Killing vector.
The transformation properties of the annihilation (or creation) operators $\delta_{\xi}\hat{a}_{\ell m}$ are completely determined from the transformation properties of the Bunch--Davies modes. To be specific, using eq. \cref{annihilation operator in terms of KG inn prod ds-d}, we find
\begin{align}
    \delta_{\xi}\hat{a}_{\ell \vec{m}} = \sqrt{\kappa_D}\kip{ \phi_{\ell \vec{m}}}{ \lie_{\xi} \hat{\varphi} } =-~\sqrt{\kappa_D} \kip{\lie_{\xi}\phi_{\ell \vec{m}}}  {\hat{\varphi}  }~, \qquad \ell \geq 1~,
\end{align}
where we have made use of the dS invariance of the Klein--Gordon inner product \cref{dS-d invariance of KG scalar product}. As mentioned earlier, positive frequency Bunch--Davies modes transform into other positive frequency Bunch--Davies modes as 
\begin{align}
    \lie_{\xi}\phi_{\ell m} = \sum_{\ell', m'} \kappa_{\ell,m}^{\ell'\!,m'}\phi_{\ell' m'}~.
\end{align}
The coefficients $\kappa_{\ell,m}^{\ell',m'}$ are the matrix elements of the dS generator $\lie_{\xi}$.%
\footnote{For the case of the dS boost $B_{\vec{0}}$ \cref{dS boost specific}, the explicit form of the coefficients $\kappa_{\ell,\vec{m}}^{\ell'\!,\vec{m}'}$ is given in eqs. \cref{dS-d transf of normalised modes ell>=2,eq:coeff-A-lm,dS-d transf of nrmlsd modes ell=1,eq:tildeA1m,eq:c0,eq:dS-d transf of constant zero-mode,eq:dS-d transf of time-dep zero-mode}.}
Thus, annihilation operators ($\ell \geq 1$) transform into other annihilation operators, as
\begin{align}
    \delta_{\xi}\hat{a}_{\ell m} =- \sum_{\ell '  m'} \left(\kappa_{\ell,m}^{\ell'\!,m'}\right)^{*}~\hat{a}_{\ell' m'},
\end{align}
and it is clear that the transformation $\delta_{\xi}$ is anti-hermitian.
Then, it is easy to show that the oscillator dS charges \cref{dS charge osillator in terms of KG general} are expressed as
    \begin{align}\label{dS-d charge osillator in terms of KG explicit}
 \hat{Q}^{\ds}_{(\text{osc})}(\xi)
       &= -{i} ~~\sum_{\ell = 1}\sum_{\vec{m}} ~\hat{a}_{\ell \vec{m}}^{\dagger}~\delta_{\xi}\hat{a}_{\ell \vec{m}} ,
\end{align}
and they are hermitian.
They generate the dS transformations
\begin{align}
   \comm{\hat{a}_{\ell \vec{m}}}{\hat{Q}^{\ds}_{(\t{osc})}(\xi)} &= 
   -\ii\,\delta_{\xi} \hat{a}_{\ell \vec{m}}\\
   \comm{\hat{a}^{\dagger}_{\ell m}}{\hat{Q}^{\ds}_{(\t{osc})}(\xi)} &= -\ii\,\delta_{\xi}\hat{a}^\dagger_{\ell \vec{m}}~.
\end{align}
It follows that the oscillator single-particle Hilbert space 
\begin{equation} 
\cH_{\t{osc},1} = \spn{\hat{a}^{\dagger}_{\ell \vec{m}}\ket{0}}_{\ell \geq 1}
\end{equation}
furnishes the same $\SO(D,1)$ UIR as the one realised on the quotient space of modes $  V_\t{BD}  =     \spn{\phi_{\ell\vec{m}}}_{\ell\geq0}\big/\spn{1}$. Note that the annihilation (creation) operator $\hat{a}_{1,\vec{0}}$  ($\hat{a}^{\dagger}_{1,\vec{0}}$) does not transform into any of the operators of the zero-mode ($\ell =0$)  sector despite the fact that the mode $\phi_{1, \vec{0}}$ transforms into the constant mode ``$1$'' as shown in \cref{dS-d transf of nrmlsd modes ell=1}.  In particular, we find
\begin{align}
\delta_{B_{\vec{0}}}  \hat{a}_{1,\vec{0}} = - \sqrt{\kappa_D} \kip{\lie_{B_{\vec{0}}}\phi_{1,\vec{0}}}{\hat{\varphi}} = -\frac{i}{2} \cA_{1,\vec{0}}\,\hat{a}_{2,\hat{0}}~, 
\end{align}
where we used \cref{dS-d transf of nrmlsd modes ell=1}, as well as $\kip{1}{\hat{\varphi}}=0$. 

\section{Various limits of the compact scalar}\label{app:limits}

In this short section, we point out how various kinds of minimally coupled massless scalars that appear in the literature can be obtained as limits of the compact minimally coupled massless scalar studied here. First let us rescale the compact scalar $\phi$ as $\Phi = f \phi$. The identification condition of the rescaled field is:
\begin{align}\label{eq:compactification-condition-appendix}
    \Phi(x) \sim \Phi(x) + 2\pi f~. 
\end{align}
The global $\ds_D$ Hilbert space of the field $\Phi$ is \cref{eq:compact-scalar-hilbert-space-dsd}:
\begin{align}
    \cH_f = L^2\qty(\S^1_f) \otimes \cF_\t{osc}~,
\end{align}
where the quantum mechanical $\S^1$ has now radius $f$ and we remind the reader that 
$\cF_\t{osc}$ is the Fock space over the single-particle oscillator Hilbert space 
\begin{align}
V_\t{BD} = \spn{\phi_{\ell \vec{m}}}_{\ell\geq 1}\Big/\spn{1}~,
\end{align} 
as in \cref{eq:VBD}. There are three interesting cases for the radius. 

\begin{itemize}
    \item[] \textbf{Case 1: $f\to \infty$.} In this case, the identification condition has no effect as the field is identified with itself infinitely far away. In other words, this is a decompactification limit wherein the scalar becomes non-compact. The zero-mode Hilbert space becomes $L^2(\R)$, and the total Hilbert space reproduces that of \cite{Kirsten:1993ug}:
    \begin{align}
        \cH_{f\to \infty} = L^2(\R)\otimes\cF_\t{osc}~. 
    \end{align}
    
    \item[] \textbf{Case 2: $f=1$.} This represents any intermediate value of $f$ (other than $0$ or $\infty$). The underlying theory is the compact minimally coupled massless scalar. See the main text. 
    
    \item[] \textbf{Case 3: $f\to 0$.} This is an interesting case where the identification condition completely collapses the zero mode. In particular the zero mode, which is the only mode affected by \cref{eq:compactification-condition-appendix}, is identified with itself shifted by an infinitesimal constant. This is equivalent to \emph{gauging} the shift symmetry of the scalar field. See for instance \cite{Anninos:2023lin}, when $D=2$. The resulting Hilbert space completely forgets the quantum mechanical zero-mode and is simply the Fock space:
    \begin{align}
        \cH_{f\to 0} = \cF_\t{osc}~. 
    \end{align}
    The quotienting procedure in \cref{eq:VBD} by $\spn{1}$ corresponds exactly to the gauged shift symmetry. The bosonic Fock space $\cF_\t{osc}$ is built using the standard  oscillator single-particle Hilbert space associated with de~Sitter representation theory; see \cref{Subsec:UIRs-BD-modes}.
    
\end{itemize}

\printbibliography

\end{document}